\documentclass{IEEEtran}
\usepackage{graphicx}        
\usepackage{subfigure}
\usepackage{amsmath}
\usepackage{balance}
\usepackage{float}
\usepackage{color}
\usepackage{url}
\usepackage{array}
\newcolumntype{L}{>{\centering\arraybackslash}m{3cm}}

\begin{document}
%
\title{A Survey on Security and Privacy Issues of Bitcoin}

\author{Mauro~Conti,~\IEEEmembership{Senior~Member,~IEEE,} Sandeep~Kumar~E,~\IEEEmembership{Member,~IEEE,} Chhagan~Lal,~\IEEEmembership{Member,~IEEE,}\\ Sushmita Ruj,~\IEEEmembership{Senior~Member,~IEEE}
\thanks{Prof. Mauro~Conti, is with Department of Mathematics, University of Padua, Padua,~Italy.~e-mail:conti@math.unipd.it. The work of M. Conti was supported by
a Marie Curie Fellowship funded by the European Commission under the agreement PCIG11-GA-2012-321980. This work is also partially supported by the EU TagItSmart! Project H2020-ICT30-2015-688061, the EU-India REACH Project ICI+/2014/342-896, the TENACE PRIN Project 20103P34XC funded by the Italian MIUR, and by the projects “Tackling Mobile Malware with Innovative Machine Learning Techniques,” “Physical-Layer Security for Wireless Communication,” and “Content Centric Networking: Security and Privacy Issues” funded by the University of Padua}
\thanks{Mr. Sandeep Kumar E, is with Department of Telecommunication Engineering, Ramaiah Institute of Technology, Bengaluru,~India. ~e-mail:sandeepe31@gmail.com}%
\thanks{Dr. Chhagan~Lal, is with Department of Mathematics, University of Padua, Padua,~Italy.~e-mail:chhagan@math.unipd.it}
\thanks{Prof. Sushmita~Ruj, is with Cryptology and Security Research Unit, Computer and Communication Sciences Division, Indian Statistical Institute, India.~e-mail:sush@isical.ac.in}
}

\maketitle
\begin{abstract}
\boldmath
Bitcoin is a popular \textit{cryptocurrency} that records all transactions in a distributed append-only public ledger called \textit{blockchain}. The security of Bitcoin heavily relies on the incentive-compatible proof-of-work (PoW) based distributed consensus protocol, which is run by network nodes called \textit{miners}. In exchange for the incentive, the miners are expected to honestly maintain the blockchain. Since its launch in 2009, Bitcoin economy has grown at an enormous rate, and it is now worth about 170 billions of dollars. This exponential growth in the market value of Bitcoin motivates adversaries to exploit weaknesses for profit, and researchers to discover new vulnerabilities in the system, propose countermeasures, and predict upcoming trends. 

\par In this paper, we present a systematic survey that covers the security and privacy aspects of Bitcoin. We start by presenting an overview of the Bitcoin protocol and its major components along with their functionality and interactions within the system. We review the existing vulnerabilities in Bitcoin and its underlying major technologies such as blockchain and PoW based consensus protocol. These vulnerabilities lead to the execution of various security threats to the normal functionality of Bitcoin. We then discuss the feasibility and robustness of the state-of-the-art security solutions. Additionally, we present current privacy and anonymity considerations in Bitcoin and discuss the privacy-related threats to Bitcoin users along with the analysis of the existing privacy-preserving solutions. Finally, we summarize the critical open challenges and suggest directions for future research towards provisioning stringent security and privacy techniques for Bitcoin.
\end{abstract}


\IEEEpeerreviewmaketitle

\section{Introduction}
\label{sec:intro}

\IEEEPARstart BITCOIN 
uses peer-to-peer (P2P) technology, and it operates without any trusted third party authority that may appear as a bank, a Chartered Accountant (CA), a notary, or any other centralized service~\cite{Nakamotos2008}. In particular, an owner has full control over its bitcoins, and she could spend them anytime and anywhere without involving any centralized authority. Bitcoin design is open-source and nobody owns or controls it. Moreover, it is a cryptographically secure electronic payment system, and it enables transactions involving virtual currency in the form of digital tokens called Bitcoin coins (BTC or simply bitcoins).

\par Since its deployment in 2009, Bitcoin has attracted a lots of attention from both academia and industry. With a market capitalization of 170 billion and more than 375,000 aggregate number of confirmed transactions per day (December 2017), Bitcoin is the most successful cryptocurrency to date. Given the amount of money at stake, Bitcoin is an obvious target for adversaries. Indeed, numerous attacks have been described targeting different aspects of the system, including double spending~\cite{KarameGhassan2012}, netsplit~\cite{HeilmanEthan2015}, transaction malleability~\cite{DeckerChr2014}, networking attacks~\cite{hijackbtc2017}, or attacks targeting mining~\cite{IttayEyalS13}~\cite{Sapirshtein2017}~\cite{Nayakkk2016} and mining pools~\cite{IttayEyal14}. In~\cite{BonneauJ2015}, authors claim that ``\emph{Bitcoin works in practice and not in theory}'' due to the lack of security research to find out theoretical foundation for Bitcoin protocols. Until today, the incomplete existence of a robust theoretical base forces the security research community for dismissing the use of bitcoins. Existing security solutions for Bitcoin lacks the required measures that could ensure an adequate level of security for its users. We believe that security solutions should cover all the major protocols running critical functions in Bitcoin, such as blockchain, consensus, key management, and networking protocols. Although, the online communities have already started to use bitcoins with the belief that Bitcoin will soon take over the online trading business. For instance, ``Wiki leaks'' request its users to donate using the bitcoins. The request quote is ``\emph{Bitcoin is a secure and anonymous digital currency, bitcoins cannot be easily tracked back to you, and are safer, and are the faster alternative to other donation methods}''. Wiki leaks also support the use of Litecoin, another cryptocurrency, for the same reason~\cite{WikiLeak}. 

\par Recently, Bitcoin technology is grabbing lots of attention from government bodies due to its increasing use by the malicious users to undermine legal controls. In~\cite{Williamf2014}, authors call bitcoins ``\emph{Enigmatic and Controversial Digital Cryptocurrency}'' due to mysterious concepts underneath the Bitcoin system and severe opposition from the government. According to~\cite{bitstatus}, the current bitcoin exchange rate is approximately USD $13k$ (as of December 2017) from around 1000 dollars at the start of 2016. The major technologies such as blockchain and consensus protocols that makes the Bitcoin a huge success will now also being envisioned in various next-generation applications, including smart trading in smart grids~\cite{TtAlam2015}, Internet of Things (IoT)~\cite{yzZhang2015}~\cite{Huckle2016}, vehicular networks~\cite{ALei2017}, healthcare data management~\cite{Mettler2016}, and smart cities~\cite{kkBiswas2016}, to name a few. As the length of popularity largely depends on the amount of security built on the system which surpasses all its other benefits, we aim to investigate the associated security and privacy issues in Bitcoin and its underlying techniques.

\subsection{Contribution}
\par In this paper, we present a comprehensive survey specifically targeting the security and privacy aspects of Bitcoin and its related concepts. We discuss the state-of-the-art attack vector which includes various user security and transaction anonymity threats that limits (or threatens) the applicability (or continuity) of bitcoins in real-world applications and services. We also discuss the efficiency of various security solutions that are proposed over the years to address the existing security and privacy challenges in Bitcoin. In particular, we mainly focus on the security challenges and their countermeasures with respect to major components of Bitcoin. In addition, we discuss the issues of user privacy and transaction anonymity along with a large array of research that has been done for enabling privacy and improving anonymity in Bitcoin. 

\par In the literature,~\cite{FTschorsch2015} provides a comprehensive technical survey on decentralized digital currencies with mainly emphasizing on bitcoins. The authors explore the technical background of Bitcoin and discuss the implications of the central design decisions for various Bitcoin technologies. In~\cite{BonneauJ2015}, authors discuss various cryptocurrencies in detail and provides a preliminary overview of the advantages and disadvantages of the use of bitcoins. However, all the existing works lack a detailed survey about security and privacy aspects of Bitcoin, and are a bit outdated, given the extensive research was done in the last couple of years on security and privacy. Moreover, there are numerous papers on Bitcoin and Cryptocurrency security and privacy however, a concise survey is required for an audience who are planning to initiate research in this direction. This paper does not attempt to solve any new challenge but presents an overview and discussion of the Bitcoin security and privacy threats along with their available countermeasures. In particular, the main contributions of this survey are as follow. 

\begin{itemize}

\item We present the essential background knowledge for Bitcoin, its functionalities, and related concepts. The goal is to enable the new readers to get the required familiarity with the Bitcoin and its underlying technologies such as transactions, blockchain, and consensus protocols. This is required in order to understand, the working methodology, benefits, and challenges that are associated with the use of bitcoins.
\item We systematically present and discuss all the existing security and privacy related threats that are associated either directly or indirectly (i.e., by exploiting one of its underlying technology) with the use of bitcoins. At various levels of its overall operation, we investigate the possibilities, which includes both practical and theoretical risks that an adversary could exploit to launch an attack on the Bitcoin.
\item We discuss the efficiency and limitations of the state-of-the-art solutions that address the security threats and enables strong privacy in Bitcoin, thus we provide a holistic technical perspective on these challenges in the use of bitcoins. Finally, based on our survey, we provide the list of lessons learned, open issues, and directions for future work.

\end{itemize}

\par To the best of our knowledge, this is the first survey that discusses and highlights the impact of existing as well as possible future security and privacy threats Bitcoin and its associated technologies. The paper aims to assist the interested readers: (i) to understand the scope and impact of security and privacy challenges in Bitcoin, (ii) to estimate the possible damage caused by these threats, and (iii) to point in the direction that will possibly lead to the detection and containment of the identified threats. In particular, the goal of our research is to raise the awareness in the Bitcoin research community on the pressing requirement to prevent various attacks from disrupting the cryptocurrency. For most of the security threats discussed in this paper, we have no evidence that such attacks have already been performed on Bitcoin. However, we believe that some of the important characteristics of Bitcoin make these attacks practical and potentially highly disruptive. These characteristics include the high centralization of Bitcoin (from a mining and routing perspective), the lack of authentication and integrity checks for network nodes, and some design choices pertaining, for instance, how in the Bitcoin network a node requests a block.


\subsection{Organization}
The rest of the paper is organized as follow. In Section~\ref{intro_Bitcoin}, we present a brief overview of Bitcoin which includes the description of its major components along with their functionalities and interactions. In Section~\ref{sec:Bit_security}, we discuss a number of security threats associated with the development, implementation, and use of bitcoins. In Section~\ref{sec:Bit_defense}, we discuss the state-of-the-art proposals that either countermeasure a security threat or enhances the existing security in Bitcoin. In Section~\ref{sec:Bit_PandA}, we discuss the anonymity and privacy threats towards the use of bitcoins along with their existing solutions. We present the summary of the observations and future research directions that are learned from our survey in Section~\ref{sec:future_work}. Finally, we conclude the paper in Section~\ref{sec:conclusion}.


\section{Overview of Bitcoin}
\label{intro_Bitcoin}

Bitcoin is a decentralized electronic payment system introduced by Nakamoto~\cite{Nakamotos2008}. It is based on peer-to-peer (P2P) network and a probabilistic distributed consensus protocol. In Bitcoin, electronic payments are done by generating transactions that transfer \textit{bitcoins} among users. The destination address (also called \textit{Bitcoin address}) is generated by performing a series of irreversible cryptographic hashing operations on the user's public key. In Bitcoin, a user can have multiple addresses by generating multiple public keys and these addresses could be associated with one or more of her wallets~\cite{wallet2014}. The private key of the user is required to spend the owned bitcoins in the form of digitally signed transactions. Using the hash of the public key as a receiving address provides the users a certain degree of anonymity, and it is recommended the practice to use different Bitcoin address for each receiving transaction. 


\par In Bitcoin, transactions are processed to verify their integrity, authenticity, and correctness by a group of resourceful network nodes called ``Miners''. In particular, instead of mining a single transaction, the miners bundle a number of transactions that are waiting for the network to get processed in a single unit called ``block''. The miner advertises a block in the whole network as soon as it completes its processing (or validation) in order to claim the mining reward. This block is then verified by the majority of miners in the network before it is successfully added in a distributed public ledger called ``blockchain''. The miner who mines a block receives a reward when the mined block is successfully added in the blockchain. We now present an overview of the major technical components and operational features that are essential for the practical realization of the Bitcoin.

\subsection{Transaction and Proof-of-Work} 
\label{sec:Bit_Tx}
\begin{figure}[h]
\centering
\includegraphics[scale = 0.40]{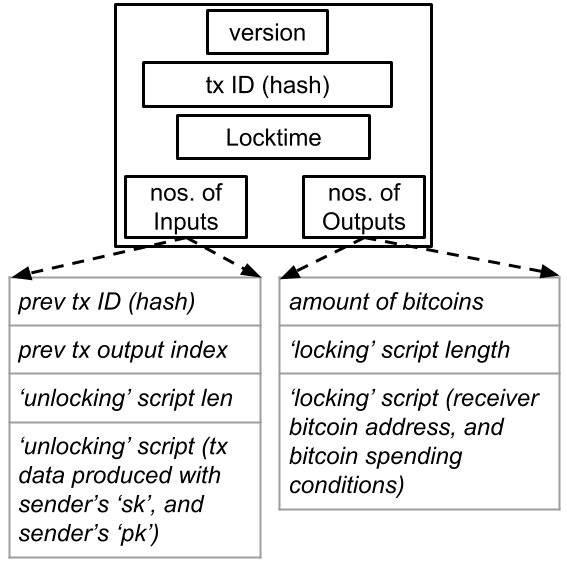}
  \caption{Bitcoin transactions}
  \label{Tx:2}
\end{figure}

Bitcoin use transactions to move coins from one user wallet to another. In particular, the coins are represented in the form of transactions, more specifically, a chain of transactions. As depicted in Figure~\ref{Tx:2}, the key fields in a transaction includes Bitcoin version, hash of the transaction, \textit{Locktime}\footnote{It indicates the earliest time or blockchain length when this transaction may be spent to the blockchain.}, one or more inputs, and one or more outputs. Every input in a transaction belongs to a particular user, and it consists of the following: (i) hash pointer to a previous transaction which serves as the identifier of the transaction that includes the output we now want to utilise as an input, (ii) an index to specific unspent previous transaction output (UTXO) that we want to spend in the current transaction, (iii) \textit{unlocking} script length, and (iv) \textit{unlocking} script (also referred to as scriptSig) which satisfies the conditions associated with the use of UTXO. While a transaction output consists of the number of bitcoins that are being transferred, \textit{locking} script length, and \textit{locking} script (also referred to as scriptPubKey) which imposes a condition that must be met before the UTXO can be spent. To authorize a transaction input, the corresponding user of the input provides the public key and the cryptographic signature generated using her private key. Multiple inputs are often listed in a transaction. All of the transaction's input values are added up, and the total (excluding transaction fee, if any) is completely used by the outputs of the transaction. In particular, when the output of a previous transaction is used as the input in a new transaction, it must be spent in its entirety. Sometimes the coin value of the output is higher than what the user wishes to pay. In this case, the sender generates a new Bitcoin address, and sends the difference back to this address. For instance, $Bob$ has 50 coins from one of its previous transaction's output, and he wants to transfer 5 coins to $Alice$ using that output as an input in a new transaction. For this purpose, $Bob$ has to create a new transaction with one input (i.e., output from its previous transaction) and two or more outputs. In the outputs, one output will show that 5 coins are transferred to Alice, and other output(s) will show transfer of the remaining coins to one (or more) wallet(s) owned by $Bob$. With this approach, the Bitcoin achieves two goals: (i) it implements the idea of $change$, and (ii) one can easily identify the unspent coins or balance of a user by only looking the outputs from its previous transactions. An output in a transaction specifies the number of coins being transferred along with the Bitcoin address of the new owner. These inputs and outputs are managed using a Forth-like scripting language which dictates the essential conditions to claim the coins. The dominant script in today's market is the ``Pay-to-PubKeyHash'' (P2PKH) which requires only one signature from the owner to authorize a payment. While the other script is called ``Pay-to-ScriptHash'' (P2SH)~\cite{GgAndresen12}, which is typically used as multi-signature addresses, but it also enables a variety of transaction types and supports future developments.

\par Unlike central bank in which all the transactions are verified, processed, and recorded in a centralized private ledger, in Bitcoin every user acts as a bank and keep a copy of this ledger. In Bitcoin, the role of the distributed ledger is played by the so-called \textit{blockchain}. However, storing multiple copies of the blockchain in the network adds new vulnerabilities in the system such as keeping the global view of the blockchain consistent. For instance, a user (say $Alice$) could generate two different transactions simultaneously using the same set of coins to two different receivers (say, $Bob$ and $Carol$). This type of malicious behavior by a user is termed as \textit{double spending}. If both the receiver processes the transaction independently based on their local view of the blockchain, and the transaction verification is successful, this leaves the blockchain into an inconsistent state. The main requirements to avoid the above problem is two-folded: (i) distribute the transaction verification process to ensure the correctness of the transaction, and (ii) everyone in the network should know quickly about a successfully processed transaction to ensure the consistent state of the blockchain. To fulfill the aforementioned requirements, Bitcoin uses the concept of \textit{Proof-of-Work} (PoW) and a probabilistic distributed \textit{consensus protocol}.

\par The distributed transaction verification process ensures that a majority of miners will verify the legitimacy of a transaction before it is added in the blockchain. In this way, whenever the blockchain goes into an inconsistent state, all the nodes update their local copy of blockchain with the state on which a majority of miners agree, in this way the correct state of the blockchain is obtained by election. However, this scheme is vulnerable to the sybil attacks~\cite{Douceur2002}. With sybil attack, a miner creates multiple virtual nodes in the network and these nodes could disrupt the election process by injecting false information in the network such as voting positive for a faulty transaction.Bitcoin counters the sybil attacks by making use of PoW based consensus model, in which to verify a transaction the miners have to perform some sort of computational task to prove that they are not virtual entities. The PoW consists of a complex cryptographic math puzzle, similar to Adam Back's \textit{Hashcash}~\cite{hashcash}. In particular, PoW involves scanning for a value (called \textit{nonce}) that when hashed, such as with SHA-256, the resulting hash begins with a number of zeros. The average work required is exponential to the number of zeros in the correct hash however, the verification process consists of a single step, i.e., by executing a single hash. In this way, PoW imposes a high level of computational cost on the transaction verification process, and the verification will be dependent on the computing power of a miner instead of the number of (possibly virtual) identities. The main idea is that it is much harder to fake the computing resources than it is to perform a sybil attack in the network.



\par In practice, the miners do not mine individual transactions instead, they collect pending transactions to form a \textit{block}. The miners mine a block by calculating the hash of that block along with a varying nonce. The nonce is varied until the resultant hash value becomes lower or equal to a given \textit{target} value. The \textit{target} is a 256-bit number that all miners share. Calculating the desired hash value is computationally difficult. For hashing, Bitcoin uses SHA-256 hash function~\cite{DEastlake2011}. Unless the cryptographic hash function finds the required hash value, the only option is to try different nonces until a solution (a hash value lower than the target) is discovered. Consequently, the difficulty of the puzzle depends on the target value, i.e., lower the target, the fewer solutions exist, hence more difficult the hash calculation becomes. Once a miner calculates the correct hash value for a block, it immediately broadcast the block in the network along with the calculated hash value and nonce, and it also appends the block in its private blockchain. The rest of the miners when receiving a mined block can quickly verify its correctness by comparing the hash value given in the received block with the \textit{target} value. The miners will also update their local blockchain by adding the newly mined block. 

\par Once a block is successfully added in the blockchain (i.e., a majority of miners consider the block valid), the miner who first solved the PoW will be rewarded (as of May 2017, $12.5$ BTCs) with a set of newly generated coins. This reward halves every 210,000 blocks. In particular, these mining rewards are not really received from anyone because there is no central authority that would be able to do this. In Bitcoin, rewards are part of the block generation process, in which a miner inserts a \textit{reward generating transaction} (or a coinbase transaction) for its own Bitcoin address, and it is always the first transaction appearing in every block. If the mined block is validated and accepted by the peers, then this inserted transaction becomes valid and the miner receives the rewarded bitcoins.

\par Apart from the mining reward, for every successful addition of a transaction in the blockchain, the miner will also receive an amount called \textit{transaction fee}, which is equivalent to the amount remaining when the value of all outputs in a transaction is subtracted from all its inputs~\cite{KKaskaloglu2014}. As the mining reward keep on decreasing with time and the number of transactions is rapidly increasing in the network, the transaction fee takes a major role for how fast a transaction is to be included in the blockchain. The Bitcoin never mandates transaction fee and it is only specified by the owner(s) of a transaction, and it is different for each transaction. A transaction with low transaction fee could suffer from the \textit{starvation} problem, i.e., denied service for a long time, if the miners are busy processing the transactions with a higher transaction fee. 

\par All the miners race to calculate the correct hash value for a block by performing the PoW, so that they can collect the corresponding reward. The chance of being the first to solve the puzzle is higher for the miners who own or control more number of computing resources. By this rule, a miner with higher computing resources can always increase her chances to win the reward. To enforce reasonable waiting time for the block validation and generation, the $target$ value is adjusted after every 2,016 blocks. This adjustment of the $target$ also helps in keeping per block verification time to approximately 10 minutes. It further effects the new bitcoins generation rate in the Bitcoin because keeping the block verification time near to 10 minutes implies that only $12.5$ new coins can be added in the network per 10 minutes. In~\cite{KraftDaniel2016}, authors propose an equation to calculate the new $target$ value for the Bitcoin. The new target is given by the following Equation.
\begin{equation}\label{eq:target}
{
T= T_{prev}*\frac{T_{actual}}{2016*10min}.
}
\end{equation}

Here, $T_{prev}$ is the old $target$ value, and $T_{actual}$ is the time period that the Bitcoin network took to generate the last 2,016 blocks.   

\begin{figure*}[h]
\centering
\includegraphics[scale = 0.35]{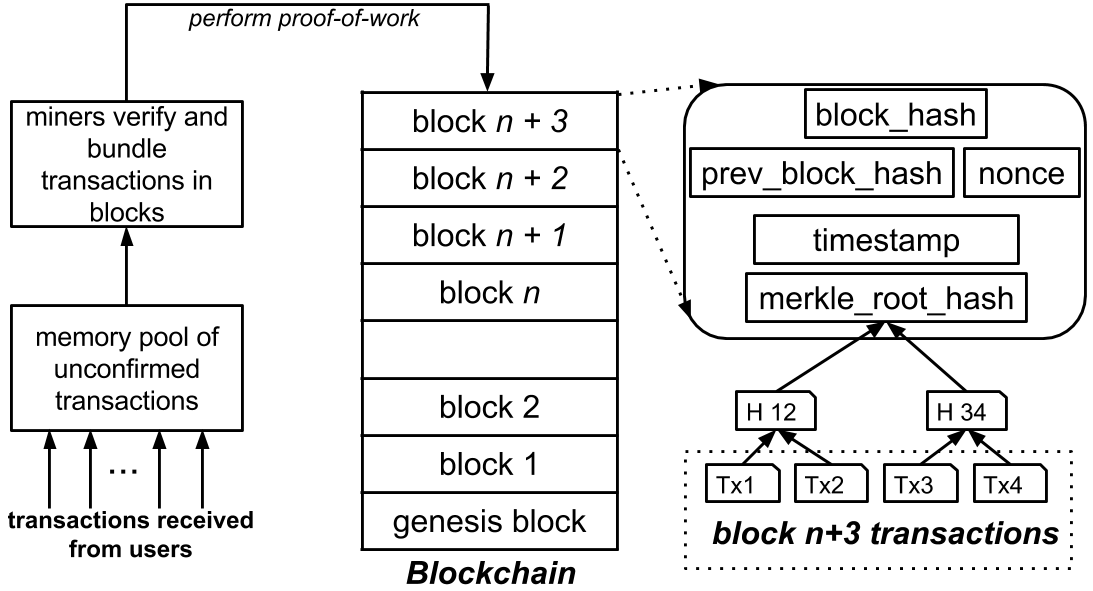}
  \caption{Creation and addition of blocks in blockchain}
  \label{block_Chain:4}
\end{figure*}

\begin{figure*}[h]
\centering
\includegraphics[scale = 0.37]{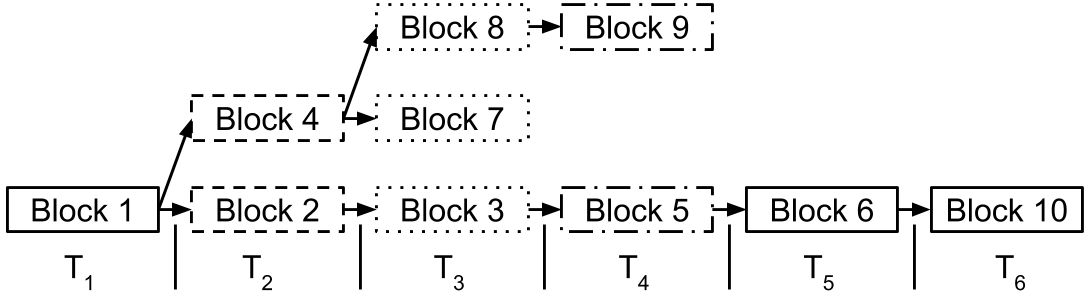}
  \caption{Blockchain consensus model}
  \label{fig:forks}
\end{figure*}
\subsection{Blockchain and Mining} 
\label{sec:Bit_mine}

\par The $blockchain$ is a public append-only link-list based data structure which stores the entire network's transaction history in terms of $blocks$. In each block, the transactions are stored using Merkle Tree~\cite{Merkle1988}, and a relatively secure time-stamp and a hash of the previous block is also stored. Figure~\ref{block_Chain:4} shows the working methodology that is being in use for creating and maintaining the Bitcoin's blockchain. To successfully add a new block in the blockchain, the miners need to verify (mine) a block by solving a computationally difficult PoW puzzle. One can traverse the blockchain in order to determine the ownership of each bitcoin because the blocks are stored in an ordered form. Also, tempering within a block is not possible as it would change the hash of the block. In particular, if a transaction in a block is tampered with, the hash value of that block changes, this, in turn, changes the subsequent blocks because each block contains the hash of the previous block. The blockchain constantly grows in length due to the continuous mining process in the network. The process of adding a new block is as follows: (i) once a miner determines a valid hash value (i.e., a hash equal or lower than target) for a block, it adds the block in her local blockchain and broadcast the solution, and (ii) upon receiving a solution for a valid block, the miners will quickly check for its validity, if the solution is correct the miners update their local copy of blockchain else discard the block. 

\par Due to the distributed nature of the block validation process, it is possible that two valid solutions are found approximately at the same time or distribution of a verified block is delayed due to network latency, this results in valid blockchain $forks$ of equal length. The forks are undesirable as the miners need to keep a global state of the blockchain, consisting of the totally ordered set of transactions. However, when multiple forks exist, the miners are free to choose a fork and continue to mine on top of it. Now that the network is having multiple forks and miners are extending different but valid versions of the blockchain based on their local view, a time will come due to the random nature of PoW where miners operating on one fork will broadcast a valid block before the others. Due to this, a longer version of the blockchain now exists in the network, and all the miners will start adding their following blocks on top of this longer blockchain. The aforementioned behavior of blockchain is shown in Figure~\ref{fig:forks}. 

\begin{figure*}[ht!]
\centering
\includegraphics[scale = 0.40]{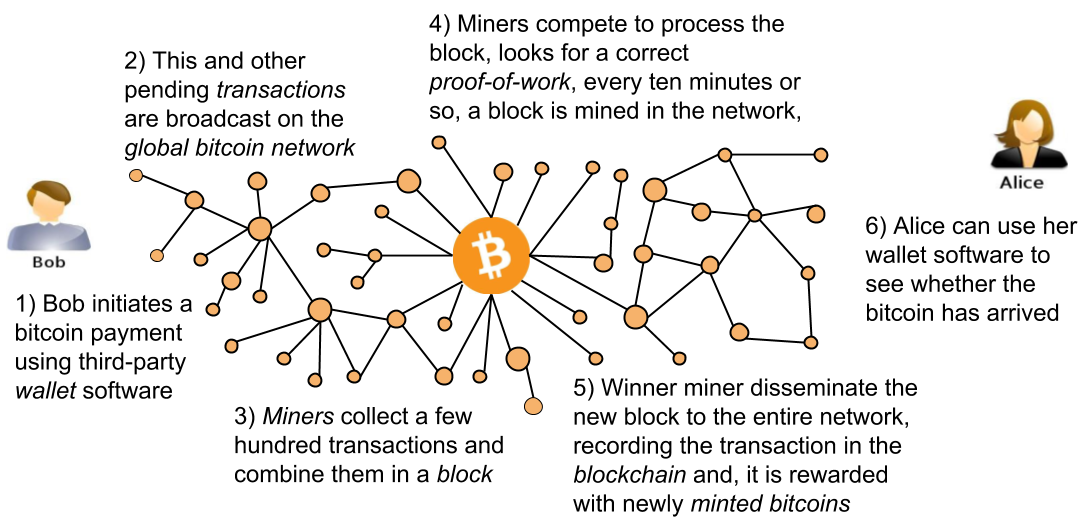}
  \caption{Bitcoin transaction processing steps}
  \label{BTC_work:5}
\end{figure*}

\par The presence of blockchain forks in Bitcoin could be exploited by a malicious miner to gain profits or to disturb the normal functioning of the Bitcoin. In particular, a resourceful miner (or mining pool) could force a blockchain fork in the network by privately mining on it. Once the malicious miner sees that the length of the public blockchain is catching up fast with her private chain, the miner broadcast her blockchain in the network, and due to its longer length, all the other miners will start mining on top of it. In this process, all the mined (i.e., valid) blocks on the other parallel blockchain get discarded which makes the efforts of the genuine miners useless. In Section~\ref{sec:Bit_security}, we will discuss an array of attacks on Bitcoin that exploits the forking nature of Bitcoin blockchain.              

\par In general, the security in Bitcoin is on the assumption that the honest players control a majority of the computing resources. The main driving factor for miners to honestly verify a block is the reward (i.e., 12.5 BTCs) that they receive upon every successful block addition in the blockchain. As mentioned before that to verify a block, the miners need to solve the associated hard crypto-puzzle. The probability of solving the crypto-puzzle is proportional to a number of computing resources used. As per~\cite{Meni2011}, a single home miner which uses a dedicated Application-Specific Integrated Circuit (ASIC) for mining will unlikely verify a single block in years. For this reason, miners mine in the form of the so-called \textit{mining pools}. All miners that are associated with a pool works collectively to mine a particular block under the control of a pool manager. Upon successful mining, the manager distributes the reward among all the associated miners proportional to the resources expended by each miner. A detailed discussion of different pooled mining approaches and their reward systems is given in~\cite{Laszka2015}~\cite{Schrijvers2017}.   

\par For the better understanding how a transaction is being processed in the Bitcoin, please refer to Figure~\ref{BTC_work:5}. Assume that $Bob$ wants to transfer 5 bitcoins to $Alice$. In order to pay to $Alice$, $Bob$ needs a device such as a smartphone, tablet, or laptop that runs the Bitcoin full or lightweight client-side software, and two pieces of information which include $Bob's$ private key and $Alice's$ Bitcoin address. Any user in the network can send money to a Bitcoin address, but only a unique signature generated using the private key can release bitcoins from the account. Bob uses a cryptographic key to digitally sign off on the transaction, proving that he owns those coins. When $Bob$ broadcast a transaction in the network, an alert is sent to all the miners in the network informing them about this new transaction. The miners check that the digital signatures are correct, and $Bob$ has enough bitcoins to complete the transactions. Additionally, miners race to bundle all the pending transactions (including $bob's$) in the network and mine the resulting block by varying the nonce. In particular, the miners create a hash of the block, and if the hash does not begin with a particular number of zeros, the hash function is rerun using a new random number (i.e., the nonce). The required hash value must have a certain but arbitrary number of zeros at the beginning. It is unpredictable which nonce will generate the required hash with a correct number of zeros, so the miners have to keep trying by using different nonces to find the desired hash value. When the miner finds a hash value with the correct number of zeros (i.e., the discovered value is lower than \textit{target} value), the discovery is announced in the network, and both the $Bob$ and the $Alice$ will also receive a confirmation about the successful transaction. Other miners communicate their acceptance, and they turn their attention to discover the next block in the network. However, a successful transaction could be discarded or deemed invalid at latter period of time, if it is unable to stay in the blockchain due to reasons, such as existence of multiple forks, majority of miners does not agree to consider the block containing this transaction as a valid block, a double spending attack is detected, to name a few.

\par The Bitcoin protocol rewards the winning miner with the set of newly minted bitcoins as $incentive$, and the hashed block is published in the public ledger. Once $Bob's$ transaction has been added in the blockchain, he and $Alice$ each receive the first confirmation stating that the Bitcoin has been signed over to $Alice$. In terms of transaction time, it depends on the current network load and the transaction fee included in the transaction by $Bob$, but at the minimum, it would be around 10 minutes. However, receiving the first confirmation does not mean that the transaction is processed successfully, and it cannot be invalidated at a latter point in time. In particular, it has been recommended by the Bitcoin community that after a block is mined it should receive enough consecutive block confirmations (currently 6 confirmations) before it is considered as a valid transaction.  

\subsection{Consensus Protocol} 
\label{sec:consensus}

Bitcoin blockchain is a decentralized system, thus it does not require authorization from any trusted third party (TTP) to process the transactions. In particular, the nodes communicate over a network and collaboratively construct the blockchain without relying on a central authority. However, individual nodes might crash, behave maliciously, act against the common goal, or the network communication may become interrupted. For delivering a continuous service, the nodes, therefore, run a fault-tolerant consensus protocol to ensure that they all agree on the order in which entries are appended to the blockchain. To add a new block in the blockchain, every miner must follow a set of rules specified in the consensus protocol. Bitcoin achieves the distributed consensus by using PoW based consensus algorithm. This algorithm imposes the following major rules: (i) input and output values are rational, (ii) transactions only spend unspent outputs, (iii) all inputs being spent have valid signatures, (iv) no coinbase\footnote{A coinbase transaction is a unique type of bitcoin transaction that can only be created by a miner.} transaction outputs were spent within 100 blocks of their creation, and (v) no transactions spend inputs with a locktime before the block in which they are confirmed. Generally, a blockchain based system such as Bitcoin is considered as secure and robust as its consensus model. 

\par In the PoW based consensus algorithm, the participants require no authentication to join the network, which makes the Bitcoin consensus model extremely scalable in terms of supporting thousands of network nodes. However, PoW based consensus is vulnerable to ``51\%'' attacks, in which an adversary has control over 51\% of the mining power (i.e. hashrate) in the network, hence it can write its own blocks or fork the blockchain that at a later point converges with the main blockchain. This behavior of adversary helps her to perform several other types of attacks in the Bitcoin, which includes double spending, eclipse, and denial-of-service. In particular, 51\% attack drives away the honest miners from the mining process, thus weakens the consensus protocol which poses a threat to Bitcoin security and robustness. One way to achieve the 51\% attack in Bitcoin system is to incentivize (or bribe) the honest miners to join the attackers' coalition.

\par Along with the various security attacks (please refer to tables~\ref{table:1} and~\ref{table:2}), the effectiveness of a consensus protocol also depends on the performance and stability of the network. For instance, an increase in the latency between the validation of a block and its receipt by all other miners increases the possibility of a temporary blockchain fork. Although, due to the PoW model eventual consistency in the blockchain will be reached despite the temporary forks however, it results in longer transaction confirmation times. Today the Bitcoin network is restricted to a sustained rate of 7 transactions per section (tps) due to the Bitcoin protocol restricting block sizes to 1MB. This is very slow when considered the high processing speed of MasterCard or VISA’s, i.e., millions of tps. Therefore, it is important for Bitcoin to have a broadcast network which is not only decentralized but it also provides low latency, and it is difficult to deliberately censor or delay messages. The PoW based consensus algorithm also wastes a lot of energy in hash computations during the mining process. However, it facilitates high scalability in terms of nodes participating in the network and operates completely in a decentralized fashion.
\par Bitcoin consensus algorithm has been its most widely debated component in the Bitcoin research community. This is because the consensus algorithm rises: (i) open questions about the Bitcoin stability~\cite{BonneauJ2015}; (ii) concerns about the performance and scalability of the protocol~\cite{Sompolinsky2015}; and (iii) concerns for computational resource wastage~\cite{LLuu2015}. In particular, the PoW consensus model used by Bitcoin blockchain is very inefficient in terms of power consumption and the overall generation time of new blocks. Hence, to overcome or limit some of the aforementioned disadvantages of PoW, various other consensus protocols such as Proof-of-Stake (PoS)~\cite{Nadal2012}, Proof of Elapsed Time (PoET), Proof of Authority (PoA), Practical byzantine fault tolerance (PBFT)~\cite{Castroo2002}, Federated Byzantine Fault Tolerance (FBFT), Proof of Storage~\cite{Millerre2014}~\cite{Sengupta2016}, to name a few are designed. The most obvious difference between these consensus protocols and PoW is that each of these alternative protocols the consensus is driven at the expense of internal resources (e.g., coins or reputation) instead of external resources (e.g., electricity). This creates an entirely different set of incentives for (and trust in) network nodes (i.e., miners), which drastically changes the network security model. Detailed discussions on these alternative consensus protocols are out of the scope of our survey, hence we direct interested users to~\cite{Bentov2016}~\cite{FTschorsch2015}~\cite{Garay2017}~\cite{Min2016}~\cite{SheharBano2017}.




\subsection{Networking Infrastructure} 
\label{sec:networks}
Bitcoin uses an unstructured peer-to-peer (P2P) network based on unencrypted persistent TCP connections as its foundational communication structure. In general, unstructured overlays are easily constructed and robust against highly dynamic network topologies, i.e., against frequently joining and leaving peers. These type of networks are best suited for Bitcoin as the aim is to distribute information as fast as possible to reach consensus on the blockchain. However, experimenting with the Bitcoin network/protocol poses a challenge. By now, there are a few possibilities to approach this task. One way is to connect to the mainnet, i.e., the live Bitcoin network, or the testnet. Another way is to use the simulation environments such as Shadow~\cite{Andreww2015} event discrete simulator, which aims at simulating large-scale Bitcoin networks, while keeping full control over all components.

\par Bitcoin nodes maintain a list of IP addresses of potential peers, and the list is bootstrapped via a DNS server, and additional addresses are exchanged between peers. Each peer aims to maintain a minimum of 8 unencrypted TCP connections in the overlay, i.e, the peer actively tries to establish additional connections if the current number of connections is lower than 8. The number of eight connections can be significantly exceeded if incoming connections are accepted by a Bitcoin peer upto a maximum of 125 connections at a time. By default, peers listen on port 8333 for inbound connections. When peers establish a new connection, they perform an application layer handshake, consisting of version and verack messages. The messages include a timestamp for time synchronization, IP addresses, and the protocol version. A node selects its peers in a random fashion and it selects a new set of peers after a fixed amount of time. This is done to minimize the possibility and effects of \textit{netsplit} attack, in which an attacker creates an inconsistent view of the network (and the blockchain) at the attacked node. Since Bitcoin version 0.7, IPv6 is supported. In order to detect when peers have left, Bitcoin uses a softstate approach. If 30 minutes have been passed since messages were last exchanged between neighbors, peers will transmit a hello message to keep the connection alive.

\par Miners continually listen to new block announcements which are sent via $INV$ messages containing the hash of the mined block. If a miner discovers that it does not hold a newly announced block, it transmits a $GETDATA$ message to one of its neighbor. The neighbor then respond by sending the requested information in a $BLOCK$ message. In case the requested block do not arrive within 20 minutes, the miner trigger the disconnection of that particular neighbor and request the same information from another neighbor. The propagation of transactions  occur in a sequence given as $INV$, $GETDATA$, and $TX$ messages, in which nodes announce, request, and share transactions that have not yet been included in the blockchain.

\par In order to form the distributed consensus, newly discovered transactions and blocks are propagated (through flooding) in the whole network. Miners store new transactions for the mining purposes, but after some time remove them if they do not make it on the blockchain. It is the responsibility of the transaction originator that the transaction is received by all the peers in the network. To this end, the originator might need to rebroadcast the transaction if it did not get into the blockchain in first attempt. This is to ensure that the transaction gets considered in the next block. An adversary could introduce delay in the propagation of both, new transactions and mined block, for the purpose of launching the double spend and netsplit attacks. As shown in~\cite{GervaisArt2015}, the propagation time can even be further extended under reasonable circumstances. Authors in~\cite{hijackbtc2017} presents a taxonomy of routing attacks and their impact on Bitcoin, considering both small-scale attacks, targeting individual nodes, and large-scale attacks, targeting the network as a whole. By isolating parts of the network or delaying block propagation, adversaries could cause significant amount of mining power to be wasted, leading to revenue losses and exposing the network to a wide range of exploits such as double spending.

\par The use of an unstructured P2P network in Bitcoin enables the required rapid distribution of information in every part of the network. The security of Bitcoin heavily depends on the global consistent state of blockchain which relies on the efficiency of its PoW based consensus protocol. The variations in the propagation mechanisms could adversely affect the consensus protocol. The presence of inconsistent blockchain states, if exploited correctly could lead to a successful double spending. To this end, it is essential that the Bitcoin network should remains scalable in terms of network bandwidth, network size, and storage requirements because this will facilitate the increase in number of honest miners in the network, which will strengthen the consensus protocol. In Bitcoin, full nodes download and verify all blocks starting from the genesis block because it is the most secure way. Full nodes participate in the P2P network and help to propagate information, although its not mandatory to do so. Alternatively, the thin clients use the simplified payment verification (SPV) to perform Bitcoin transactions. The SPV is a method used by Bitcoin thin client for verifying if particular transactions are included in a block without downloading the entire block. However, the use of SPV costs the thin clients because it introduces weaknesses such as Denial of Service (DoS) and privacy leakage for the thin client. In particular, the general scalability issues of unstructured overlays combined with the issues induced by the Bitcoin protocol itself remains in the system. Many of the results suggest that scalability remains an open problem~\cite{Courtoise2014} and it is hard to keep the fully decentralized network in future~\cite{KrollTheEO2013}~\cite{Gervais14}.

\subsection{Benefits and Challenges}
\label{sec:Bit_proscons}
\par Same as any other emerging technology, use of Bitcoin comes with certain benefits and challenges, and various types of risks are associated with its use. It is believed\footnote{As some of these benefits and challenges are not entirely true at all the times, for instance, Bitcoin transactions are not fully anonymous and the privacy of Bitcoin users could be threatened.} that Bitcoin has the following benefits and challenges. \\

\emph{\textbf{Benefits -}}
\begin{itemize}

  \item \textit{No Third-Party Seizure:} No central authority can manipulate or seize the currency since every currency transfer happens peer-to-peer just like hard cash. In particular, bitcoins are yours and only yours, and the central authority can’t take your cryptocurrency, because it does not print it, own it, and control it correspondingly.
 
  \item \textit{Anonymity and transparency:} Unless Bitcoin users publicize their wallet addresses publicly, it is extremely hard to trace transactions back to them. However, even if the wallet addresses was publicized, a new wallet address can be easily generated. This greatly increases privacy when compared to traditional currency systems, where third parties potentially have access to personal financial data. Moreover, this high anonymity is achieved without sacrificing the system transparency as all the bitcoin transactions are documented in a public ledger.

 \item \textit{No taxes and lower transaction fees:} Due to its decentralized nature and user anonymity, there is no viable way to implement a Bitcoin taxation system. In the past, Bitcoin provided instant transactions at nearly no cost. Even now, Bitcoin has lower transaction costs than a credit card, Paypal, and bank transfers. However, the lower transaction fee is only beneficial in situations where the user performs a large value international transactions. This is because the average transaction fee becomes higher for very small value transfers or purchases such as paying for regular household commodities.

 \item \textit{Theft resistance:} Stealing of bitcoins is not possible until the adversary have the private keys (usually kept offline) that are associated with the user wallet. In particular, Bitcoin provides security by design, for instance, unlike with credit cards you don’t expose your secret (private key) whenever you make a transaction. Moreover, bitcoins are free from \textit{Charge-backs}, i.e., once bitcoins are sent, the transaction cannot be reversed. Since the ownership address of the sent bitcoins will be changed to the new owner, and it is impossible to revert. This ensures that there is no risk involved when receiving bitcoins.
\end{itemize}

\emph{\textbf{Challenges:}}
\begin{itemize}
  
 \item \textit{High energy consumption:} Bitcoin's blockchain uses PoW model to achieve distributed consensus in the network. Although, the use of PoW makes the mining process more resistant to various security threats such as sybil and double spending, it consumes a ridiculous amount of energy and computing resources~\cite{Fairley2017}. In particular, processing a bitcoin transaction consumes more than 5000 times as much energy as using a Visa credit card, hence innovative technologies that reduce this energy consumption are required to ensure a sustainable future for Bitcoin. Furthermore, due to the continuous increase in network load and energy consumption, the time required for transaction processing is increasing.

   \item \textit{Wallets can be lost:} Since there is no trusted third party if a uses lost the private key associated with her wallet due to a hard drive crash or a virus corrupts data or lost the device carrying the key, all the bitcoins in the wallet has been considered lost for \textit{forever}. There is nothing that can be done to recover the bitcoins, and these will be forever orphaned in the system. This can bankrupt a wealthy Bitcoin investor within seconds.
 
   \item \textit{(Facilitate) Criminal activity:} The considerable amount of anonymity provided by the Bitcoin system helps the would-be cyber criminals to perform various illicit activities such as ransomware~\cite{Liao2016}, tax evasion, underground market, and money laundering.

\end{itemize}

\par According to~\cite{MariamK2014}, the risk is the exposure to the level of danger associated with Bitcoin technology; in fact, the same can be applied to any such digital cryptocurrency. The major risks that threaten the wide usability of the Bitcoin payment systems are as follow.
\begin{itemize}
  \item \emph{Social risks:}~it includes bubble formation (i.e., risk of socio-economic relationship such as what people talk and gossip), cool factor (i.e., entering the networking without knowing the ill effects), construction of chain (i.e., risk related with the blockchain formation like hashing and mining rewards), and new coins release (i.e., on what basis the new coins to be generated, is there a need etc.).
  \item \emph{Legal risks:}~Bitcoin technology opposes rules and regulations, and hence it finds opposition from the government. This risk also includes law enforcement towards handling financial, operational, customer protection and security breaches that arise due to Bitcoin system.
  \item \emph{Economic risks:}~deflation, volatility and timing issues in finding a block which might lead the users to migrate towards other currencies that offer faster services.
    \item \emph{Technological risks:}~this includes the following, network equipment, and its loss, network with which the peers are connected and its associated parameters, threat vulnerabilities on the system, hash functions with its associated robustness factor, and software associated risks that Bitcoin system demands.
   \item \emph{Security risks:}~security is a major issue in Bitcoin system, we will discuss risks associated due to various security threats in detail in Section~\ref{sec:Bit_security}.
\end{itemize}
In~\cite{Bashirrr2016}, authors perform a survey on the people's opinion about bitcoins usage. Participants argue that the greatest barrier to the usage of bitcoins is the lack of support by higher authorities (i.e., government). Participants felt that bitcoins must be accepted as legitimate and reputable currency. Additionally, the participants expressed that the system must provide support towards transacting fearlessly without criminal exploitation. Furthermore, the Bitcoin is mainly dependent on the socio-technical actors, and the impact of their opinion on the public. Few among participants have suggested that the blockchain construction is the major cause of disruption due to its tendency to get manipulated by adversaries.
\par In~\cite{Krombholz2017}, it was stated that many Bitcoin users already lost their money due to poor usability of key management and security breaches, such as malicious exchanges and wallets. Around 22.5\% of the participants reported having lost their bitcoins due to security breaches. Also, many participants stated that for a fast flow of bitcoins in the user community, simple and impressive user interface are even more important than security. In addition, participants highlighted that the poor usability and lack of knowledge regarding the Bitcoin usage are the major contributors to the security failures.

\section{Security: Attacks on Bitcoin Systems}
\label{sec:Bit_security}

Bitcoin is the most popular cryptocurrency\footnote{www.cryptocoinsnews.com/} and has stood first in the market capital investment from day one. Since it is a decentralized model with an uncontrollable environment, hackers and thieves find cryptocurrency system an easy way to fraud the transactions. In this section, we discuss existing security threats and their countermeasures for Bitcoin and its underlying technologies. We provide a detailed discussion of potential vulnerabilities that can be found in the Bitcoin protocols as well as in the Bitcoin network, this will be done by taking a close look at the broad attack vector and their impact on the particular components in the Bitcoin. Apart from double spending, which is and will always be possible in Bitcoin, the attack space includes a range of wallet attacks (i.e., client-side security), network attacks (such as DDoS, sybil, and eclipse) and mining attacks (such as 50\%, block withholding, and bribery). Tables~\ref{table:1} and~\ref{table:2} provides a comprehensive overview of the potential security threats along with their impacts on various entities in Bitcoin and their possible solutions that exist in literature so far.

\begin{table*}[h]
\caption {Major attacks on Bitcoin system and its PoW based consensus protocol}
\centering
\scalebox{1.06}{
\begin{tabular}{|L|p{3cm}|p{2cm}|L|p{4cm}|}\hline

\textbf{Attack} & \textbf{Description} & \textbf{Primary targets} & \textbf{Adverse effects} & \textbf{Possible countermeasures} \\ \hline

\textit{Double spending or Race attack}~\cite{KarameGhassan2012} & spent the same bitcoins in multiple transactions, send two conflicting transactions in rapid succession & sellers or merchants & sellers lose their products, drive away the honest users, create blockchain forks & inserting observers in network~\cite{KarameGhassan2012}, communicating double spending alerts among peers~\cite{KarameGhassan2012}, nearby peers should notify the merchant about an ongoing double spend as soon as possible~\cite{KarameAndroulaki2015}, merchants should disable the direct incoming connections~\cite{BamertDecker2013}~\cite{LearBahack2013}\\ \hline

\textit{Finney attack}~\cite{HFinney2011} & dishonest miner broadcasts a pre-mined block for the purpose of double spending as soon as it receives product from a merchant & sellers or merchants & facilitates double spending & wait for multi-confirmations for transactions\\\hline

\textit{Brute force attack}~\cite{JHeusser2013} & privately mining on blockchain fork to perform double spending & sellers or merchants & facilitates double spending, creates large size blockchain forks & inserting observers in the network~\cite{KarameGhassan2012}, notify the merchant about an ongoing double spend~\cite{BamertDecker2013}\\\hline

\textit{Vector 76 or one-confirmation attack}~\cite{Vector672011} & combination of the double spending and the finney attack & Bitcoin exchange services & facilitates double spending of larger number of bitcoins & wait for multi-confirmations for transactions\\\hline

\textit{$> 50\%$ hashpower or Goldfinger}~\cite{KrollTheEO2013} & adversary controls more than $> 50\%$ Hashrate & Bitcoin network, miners, Bitcoin exchange centers, and users & drive away the miners working alone or within small mining pools, weakens consensus protocol, DoS & inserting observers in the network~\cite{KarameGhassan2012}, communicating double spending alerts among peers~\cite{KarameGhassan2012}, disincentivize large mining pools~\cite{IEyal2014}~\cite{MartijnBastiaan2015}, TwinsCoin~\cite{TwinsCoin2017}, PieceWork~\cite{Daiannm2017}\\\hline

\textit{Block discarding}~\cite{NicolasT2014}~\cite{LearBahack2013} or Selfish mining~\cite{IttayEyalS13} & abuses Bitcoin forking feature to derive an unfair reward & honest miners (or mining pools) & introduce race conditions by forking, waste the computational power of honest miners, with $> 50\%$ it leads to Goldfinger attack & ZeroBlock technique~\cite{ZhangRen2017}~\cite{SiamakSolatP16}, timestamp based techniques such as freshness preferred~\cite{HeilmanEthan2014}, DECOR+ protocol~\cite{DLerner2014}\\\hline

\textit{Block withholding}~\cite{Meni2011}~\cite{SBag2016} & miner in a pool submits
only PPoWs, but not FPoWs & honest miners (or mining pools) & waste resources of fellow miners and decreases the pool revenue & include only known and trusted miners in pool, dissolve and close a pool when revenue drops from expected~\cite{NicolasT2014}, cryptographic commitment schemes~\cite{SBag2016} \\\hline

\textit{fork after withholding (FAW) attack}~\cite{Kwon2017} & improves on adverse effects of selfish mining and block withholding attack & honest miners (or mining pools) & waste resources of fellow miners and decreases the pool revenue & no practical defense reported so far \\\hline

\end{tabular}}
\label{table:1}
\end{table*}

\subsection{Double Spending}
\label{sec:Bit_DB}

\begin{figure}[h]
\centering
\includegraphics[scale = 0.23]{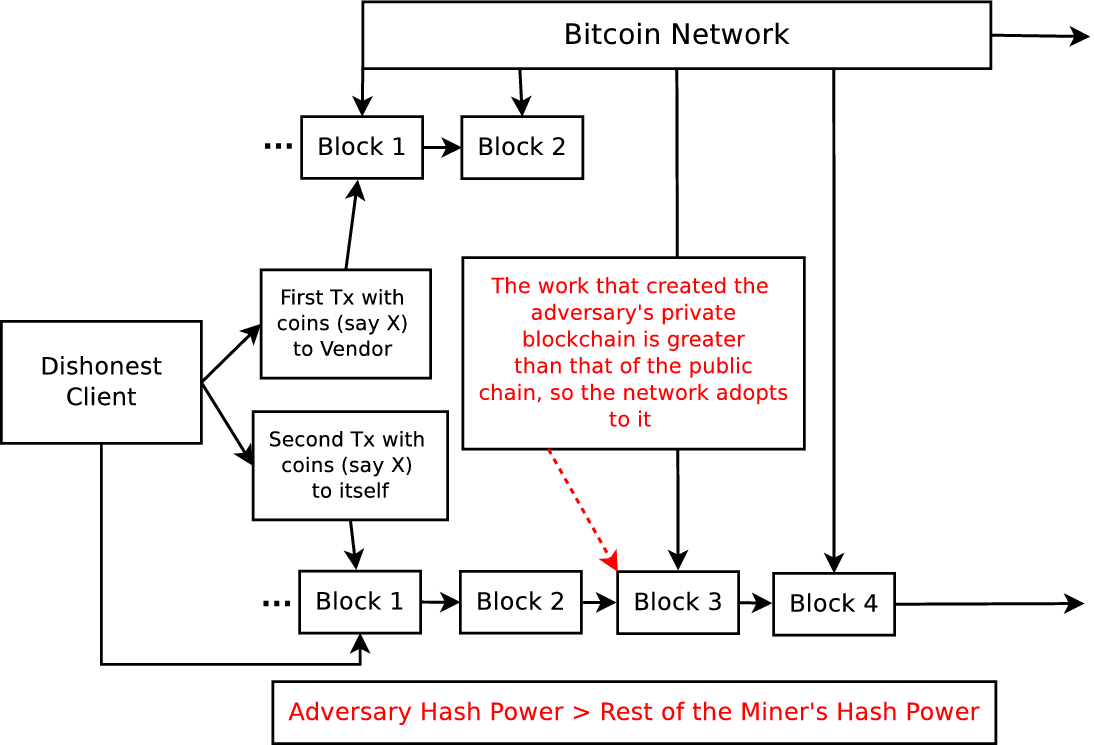}
  \caption{{Double Spending Attack}}
  \label{fig:ds}
\end{figure}

A client in the Bitcoin network achieves a double spend (i.e., send two conflicting transactions in rapid succession) if she is able to simultaneously spend the same set of bitcoins in two different transactions~\cite{KarameGhassan2012}. For instance, a dishonest client ($C_d$) creates a transaction $T_V^{C_d}$ at time $t$ using a set of bitcoins ($B_c$) with a recipient address of a vendor ($V$) to purchase some product from $V$. $C_d$ broadcast $T_V^{C_d}$ in the Bitcoin network. At time $t'$ where $t'$ $\approx$ $t$, $C_d$ create and broadcast another transaction $T_{C_d}^{C_d}$ using the same coins (i.e., $B_c$) with the recipient address of $C_d$ or a wallet which is under the control of $C_d$. In the above scenario, the double spending attack performed by $C_d$ is successful, if $C_d$ tricks the $V$ to accept $T_V^{C_d}$ (i.e., $V$ deliver the purchased products to $C_d$) but $V$ will not be able to redeem subsequently. In Bitcoin, the \textit{network of miners} verify and process all the transactions, and they ensure that only the unspent coins that are specified in previous transaction outputs can be used as input for a follow-up transaction. This rule is enforced dynamically at run-time to protect against the possible double spending in the network. The distributed time-stamping and PoW-based consensus protocol is used for orderly storage of the transactions in the blockchain. For example, when a miner receives $T_V^{C_d}$ and $T_{C_d}^{C_d}$ transactions, it will be able to identify that both the transactions are trying to use the same inputs during the transaction propagation and mining, thus it only process one of the transaction and reject the other. Figure~\ref{fig:ds} shows the working methodology of a double spending attack depicting the above explanation. 

\par Despite the use of strict ordering of transactions in the blockchain, PoW scheme, distributed time-stamping~\cite{Haber1991}, and consensus protocol~\cite{DMalkhi1988}~\cite{NSzabo1988}, double spending is still possible in Bitcoin. To perform a successful double spending attack, following requirements need to be fulfilled: (i) part of the network miners accept the transaction $T_V^{C_d}$ and the vendor ($V$) receives the confirmation from these miners, thus releases the product to dishonest client ($C_d$), (ii) at the same time, other part of the network miners accept the transaction $T_{C_d}^{C_d}$, hence lead to blockchain forks in the network, (iii) the vendor receives the confirmation of transaction $T_{C_d}^{C_d}$ after accepting the transaction $T_V^{C_d}$, thus losses the product, and (iv) a majority of miners mine on top of the blockchain which contains $T_{C_d}^{C_d}$ as a valid transaction. If the aforementioned steps took place in the given order then the dishonest client is able to perform a successful double spend. In the rest of this section, we will discuss the variants of double spending attack that are used in order to realize the aforementioned double spend requirements with varying difficulties and complexities.        

\par A form of double spending called \textit{Finney attack}~\cite{HFinney2011}, here a dishonest client ($C_d$) pre-mines (i.e., privately) a block which contains the transaction $T_{C_d}^{C_d}$, and then it creates a transaction $T_V^{C_d}$ using the same bitcoins for a vendor ($V$). The mined block is not informed to the network, and the $C_d$ waits until the transaction $T_V^{C_d}$ is accepted by the $V$. On the other hand, $V$ only accept $T_V^{C_d}$ when it receives a confirmation from miners indicating that $T_V^{C_d}$ is valid and included in the existing blockchain. Once $C_d$ receives the product from $V$, the attacker releases the pre-mined block into the network, thus creates a blockchain fork (say $B'_{fork}$) of equal length to the existing fork (say $B_{fork}$). Now, if the next mined block in the network extends $B'_{fork}$ blockchain instead of $B_{fork}$, then as per the Bitcoin protocol rules all the miners in the network will build on top of $B'_{fork}$. As the blockchain $B'_{fork}$ becomes the longest chain in the network, all the miners ignore $B_{fork}$, hence the top block on $B_{fork}$ which contains the transaction $T_V^{C_d}$ becomes invalid. This makes the transaction $T_V^{C_d}$ invalid, the client will get back her coins through transaction $T_{C_d}^{C_d}$, but resulting the $V$ losing the product. However, with \textit{Finney attack} an adversary can only perform double spending in the presence of one-confirmation vendors.  

\par To avoid the \textit{Finney attack}, the vendor should wait for multiple confirmations before releasing the product to the client. The waiting for multiple confirmations will only make the double spend for the attacker harder, but the possibility of the double spend remains. An advancement of the \textit{Finney attack} is called \textit{Brute-force attack}~\cite{JHeusser2013} in which a resourceful attacker has control over $n$ nodes in the network, and these nodes collectively work on a private mining scheme with the motive of double spend. An attacker introduces a double spend transaction in a block as in the previous case, while continuously works on the extension of a private blockchain (i.e., $B'_{fork}$). Suppose a vendor waits for $x$ confirmations before accepting a transaction, and it sends the product to the client once it receives the $x$ confirmations. Later, the attacker is able to mine the $x$ number of blocks ahead (i.e., privately) then she can release these blocks in the network, and due to its higher length than $B_{fork}$, blockchain $B'_{fork}$ will be extended by all the miners in the network. This causes the same after effects as \textit{Finney attack}, thus causing a successful double spending attack.

\par Another attack that uses the privately mined block to perform a new form of double spending attack on Bitcoin exchange networks is popularly known as \textit{Vector 76 attack}~\cite{Vector672011}. A Bitcoin exchange is a digital marketplace where traders can buy, sell or exchange bitcoins for other assets, such as fiat currencies or altcoins. In this, a dishonest client ($C_d$) withholds a pre-mined block which consists of a transaction that implements a specific deposit (i.e., deposit coins in a Bitcoin exchange). The attacker ($C_d$) waits for the next block announcement and quickly sends the pre-mined block along with the recently mined block directly to the Bitcoin exchange or towards its nearby peers with hope that the exchange and probably some of the nearby miners will consider the blockchain containing the pre-mined block (i..e, $B'_{fork}$) as the main chain. The attacker quickly sends another transaction that requests a withdrawal from the exchange of the same coins that was deposited by the attacker in its previous transaction. At this point of time, if the other fork (i.e., $B_{fork}$) which does not contain the transaction that is used by the attacker to deposit the coins survives, the deposit will become invalidated but the attacker has already performed a withdrawal by now, thus the exchanges losses the coins. 

\par Recently, authors in~\cite{DBLPNatoliG16a} proposes a new attack against the PoW-based consensus mechanism in Bitcoin called the \textit{Balance attack}. The attack consists of delaying network communications between multiple subgroups of miners with balanced hash power. The theoretical analysis provides the precise trade-off between the Bitcoin network communication delay and the mining power of the attacker(s) needed to double spend in Ethereum~\cite{GWood2015} with high probability.

\par Based on the above discussion on double spending attack and its variants, one main point that emerges is that if a miner (or mining pool) is able to mine blocks with a faster rate than the rest of the Bitcoin network, the possibility of a successful double spending attack is high. The rate of mining a block depends upon solving the associated proof-of-work, this again depends on the computing power of a miner. Apart from the computing resources, the success of double spending attack depends on other factors as well which includes network propagation delay, vendor, client, and Bitcoin exchange services connectivity or positioning in the Bitcoin network, and the number of honest miners. Clearly, as the number of confirmations for transaction increases, the possibility that it will become invalid at a later stage decreases, thus decreases the possibility of a double spend. On the other hand, with the increase in the computing resources of a miner, the probability of the success of a double spend increases. This leads to a variant of double spend attack called \textit{$>50\%$ attack} or \textit{Goldfinger attack}~\cite{KrollTheEO2013} in which more than 50\% computing resources of the network are under the control of a single miner (or mining pool). The \textit{$>50\%$ attack} is considered the worst-case scenario in the Bitcoin network because it has the power to destroy the stability of the whole network by introducing the actions such as claim all the block intensives, perform double spending, reject or include transactions as preferred, and play with the Bitcoin exchange rates. The instability in the network once started, it will further strengths the attacker's position as more and more honest miners will start leaving the network.         

\par From the above discussion on the different type of double spending attacks, we can safely conclude that one can always perform a double spend or it is not possible to entirely eliminate the risk of double spending in Bitcoin. However, performing double spending comes with a certain level of risk, for instance, the attacker might lose the reward for the withheld block if it is not included in the final public blockchain. Therefore, it is necessary to set a lower bound on the number of double spend bitcoins, and this number should compensate the risk of unsuccessful attempts of double spend. Additionally, the double spends could be recognized with the careful analysis and traversing of the blockchain, thus it might lead to blacklisting the detected peer. In Section~\ref{sec:Bit_defense_DB}, we will discuss in detail, the existing solutions and their effectiveness for detecting and preventing the double spending attacks.

\subsection{Mining Pool Attacks}
\label{sec:Bit_MPool}

Mining pools are created in order to increase the computing power which directly affects the verification time of a block, hence it increases the chances of winning the mining reward. For this purpose, in recent years, a large number of mining pools have been created, and the research in the field of miner strategies is also evolved. Generally, mining pools are governed by pool managers which forwards unsolved work units to pool members (i.e., miners). The miners generate \textit{partial proofs-of-work} (PPoWs) and \textit{full proofs-of-work (FPoWs)}, and submit them to the manager as \textit{shares}. Once a miner discovers a new block, it is submitted to the manager along with the FPoW. The manager broadcasts the block in the Bitcoin network in order to receive the mining reward. The manager distributes the reward to participating miners based on the fraction of shares contributed when compared with the other miners in the pool. Thus, participants are rewarded based on PPoWs, which have absolutely no value in the Bitcoin system. The Bitcoin network currently consists of solo miners, open pools that allow any miner to join, and closed (private) pools that require a private relationship to join.

\par
\begin{figure}[h]
\centering
\includegraphics[scale = 0.62]{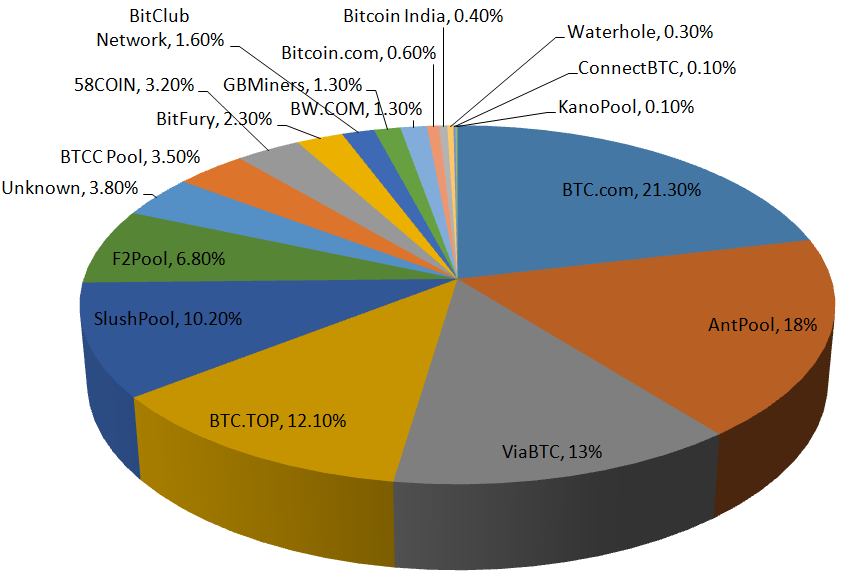}
  \caption{{Bitcoin Hashrate Distribution in Present Market}}
  \label{fig:MPool}
\end{figure}

 In recent years, the attack vector that exploits the vulnerabilities in pool based mining also increases. For instance, a group of dishonest miners could perform a set of internal and external attacks on a mining pool. Internal attacks are those in which miners act maliciously within the pool to collect more than their fair share of collective reward or disrupt the functionality of the pool to distant it from the successful mining attempts. In external attacks, miners could use their higher hash power to perform attacks such as double spending. Figure~\ref{fig:MPool} shows the market share till December 2017 of the most popular mining pools. In this section, we will discuss a set of popular internal and external attacks on the mining pools.

\par In a mining pool, the pool manager determines the amount of work done by individual pool members, by using the number of \text{shares}, a member find and submit while trying to discover a new block. The shares consist of a number of hashes of a block which are low enough to have discovered a block if the difficulty was $1$. To be considered as a share, each hash has a probability of $1/2^{32}$. Assuming correctness of the hash function used, it is impossible to find shares without doing the work required to discover new blocks or to look for blocks without finding shares along the way. Due to this, the number of shares determined by a miner is proportional, on average, to the number of hashes the miner calculated while attempting to discover a new block for the mining pool. Additionally, in~\cite{Meni2011}, the author discusses the possibility of using variable block rewards and difficulty shares as reward methods in a pool. This variability is introduced due to the following reasons; bitcoins generation per block is cut in half every 210000 blocks, and the transaction fees vary rapidly based on the currently available transactions in the network. As most of the mining pools allow any miner to join them using a public Internet interface, such pools are susceptible to various security threats. The adversaries believe that it is profitable to \textit{cannibalize} pools than mine honestly. Let's understand it with an example, suppose that an adversary has 30\% of hashrate (HR) and 1 BTC is the block mining reward (MR). If the mining pool is sharing the reward based on the invested HR then the adversary will receive 0.3 BTC for each mined block. Now adversary purchases more mining equipment, worth 1\% of current HR. With standard mining strategy, the adversary will gain an additional revenue of 0.0069 BTC for the 1\% added HR. By performing pool cannibalizing (i.e., distribute your 1\% equally among all other pools, and also withhold the valid blocks) the attacker will still receive the rewards from its pool, but it might also receive additional rewards from the other pools to which she is sharing its 1\% HR. This misbehavior will remain undetectable unless the change in reward is statistically significant.

\begin{figure}[h]
\centering
\includegraphics[scale = 0.30]{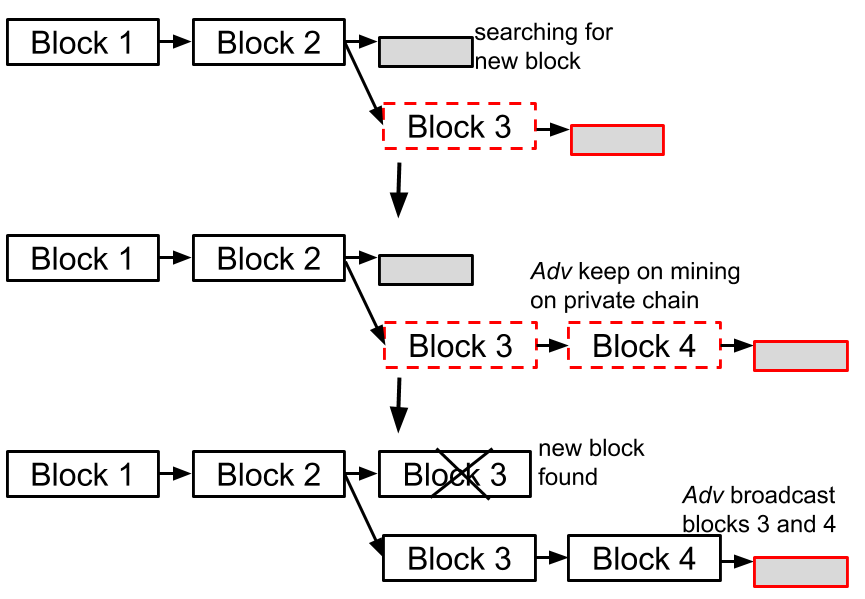}
  \caption{Selfish Mining}
  \label{fig:fish}
\end{figure}

\par In~\cite{NicolasT2014}, authors use a game theoretic approach to show that the miners could have a specific sort of subversive mining strategy called \textit{selfish mining}~\cite{IttayEyalS13} or also popularly known as \textit{block discarding attack}~\cite{LearBahack2013}~\cite{NicolasT2014}. In truth, all the miners in the Bitcoin are \textit{selfish} as they are mining for the reward that is associated with each block, but these miners are also honest and fair with respect to the rest of miners, while the \textit{selfish mining} here refers to the malicious miners only. In the selfish mining, the dishonest miner(s) perform information hiding (i.e., withhold a mined block) as well as perform its revealing in a very selective way with a two-fold motive: (i) obtain an unfair reward which is bigger than their share of computing power spent, and (ii) confuse other miners and lead them to waste their resources in a wrong direction. As it can be seen in Figure~\ref{fig:fish} that by keeping the mined block(s), the selfish miners intentionally fork the blockchain. The selfish pool keeps on mining on top of their private chain ($B'_{fork}$), while the honest miners are mining on the public chain ($B_{fork}$). If the selfish miners are able to take a greater lead on $B'_{fork}$ and they are able to keep the lead for a longer time period, their chances of gaining more reward coins as well as the wastage of honest miners resources increases. To avoid any losses, as soon as the $B_{fork}$ reaches to the length of $B'_{fork}$, the selfish miners publish their mined blocks. All the miners need to adopt to $B'_{fork}$ which now becomes $B_{fork}$ as per the longest length rule of Bitcoin protocol. The honest miners will lose their rewards for the blocks that they have mined and added to the previous public chain. The analysis in~\cite{IttayEyalS13} shows that using the selfish mining, the pool's reward exceed its share of the network's mining power. The statement still holds in cases where the network found their new block before the adversary could find a new second block. Because in such case the miner will make use of the \textit{race to propagate}, i.e., on average the attacker manages to tell 50\% of the network about her block first. Additionally, the analysis reveals that the wastage of computing resources and rewards lure honest miners toward the selfish mining pools, hence it further strengthens the attack. This continuous increase in the selfish pool's size might lead to $>50\%$ attack, and at that point, the effect of selfish mining will be disastrous.

\par Another attack much similar to the selfish mining that could be performed on a mining pool is known as \textit{Block withholding} (BWH)~\cite{Meni2011}~\cite{SBag2016}, in which a pool member never publishes a mined block in order to sabotage the pool revenue however, submit shares consists of PPoWs, but not FPoWs. In particular, in~\cite{Meni2011}, two types of block withholding scenarios are presented called ``Sabotage'' and ``Lie in wait''. In the first scenario, the adversary does not gain any bitcoins, but it just makes other pool members lose, while in the second scenario, the adversary performs a complex block concealing attack similar to the one described in the \textit{selfish mining} attack. In~\cite{Meni2011}, authors discuss a generalized version of the ``Sabotage'' attack which shows that with slight modification, it is possible for the malicious miner to also earn an additional profit in this scenario. Authors in~\cite{LLuu2015} present a game-theoretic approach to analyzing effects of block withholding attack on mining pools. The analysis shows that the attack is always well-incentivized in the long-run, but may not be so for a short duration. This implies that existing pool protocols are insecure, and if the attack is conducted systematically, Bitcoin pools could lose millions of dollars worth in just a few months. 

\par To analyze the effects of BWH on mining pools, authors in~\cite{IttayEyal14} presents \textit{The Miner’s Dilemma}, which uses an iterative game to model attack decisions. The game is played between two pools, say pool $A$ and pool $B$, and each iteration of the game is a case of the \textit{Prisoner’s Dilemma}, i.e., choose between attacking or not attacking. If pool $A$ chooses to attack pool $A$, pool $A$ gains revenue, pool $A$ loses revenue, but pool B can latter retaliate by attacking pool A and gaining more revenue. Thus, attacking is the dominant strategy in each iteration, hence if both pool A and pool B attack each other, they will be at a Nash Equilibrium. This implies that if both will earn less than they would have if neither of them attacked. However, if none of the other pools attack, a pool can increase its revenue by attacking the others. Recently, authors in~\cite{Kwon2017} propose a novel attack called a \textit{fork after withholding} (FAW) attack. Authors show that the BWH attacker’s reward is the lower bound of the FAW attacker’s, and it is usable up to four times more often per pool than in BWH attack. Moreover, the extra reward for a FAW attack when operating on multiple mining pools is around 56\% higher than BWH attack. Furthermore, the miner’s dilemma may not hold under certain circumstances, e.g., when two pools execute FAW attack, the larger pool can consistently win. More importantly, unlike selfish mining, an FAW attack is more practical to execute while using intentional forks. 

\par The \textit{Pool Hopping attack} presented in~\cite{Meni2011}~\cite{MeniRosen2013} uses the information about the number of submitted shares in the mining pool to perform the selfish mining. In this attack, the adversary performs continuous analysis of the number of shares submitted by fellow miners to the pool manager in order to discover a new block. The idea is that if already a large number of shares have been submitted and no new block has been found so far, the adversary will be getting a very small share from the reward because it will be distributed based on the shares submitted. Therefore, at some point in time, it might be more profitable for the adversary to switch to another pool or mine independently.

\par Recently, the \textit{Bribery attack} is described in~\cite{BonneauJ16}. In this, an attacker might obtain the majority of computing resources for a short duration via bribery. Authors discuss three ways to introduce bribery in the network: (i) Out-of-Band Payment, in which the adversary pays directly to the owner of the computing resources and these owners then mine blocks assigned by the adversary, (ii) Negative-Fee Mining Pool, in which the attacker forms a pool by paying higher return, and (iii) In-Band Payment via Forking, in which the attacker attempts to bribe through Bitcoin itself by creating a fork containing bribe money freely available to any miner adopting the fork. By having the majority of the hash power, the attacker could launch different attacks such as double spending and Distributed Denial-of-Service (DDoS)~\cite{NeilGandal16}. The miners that took the bribes will get benefits which will be short-lived, but these short-lived benefits might be undermined by the losses in the long run due to the presence of DDoS and Goldfinger attacks or via an exchange rate crash.    

\begin{figure}[h]
\centering
\includegraphics[scale = 0.23]{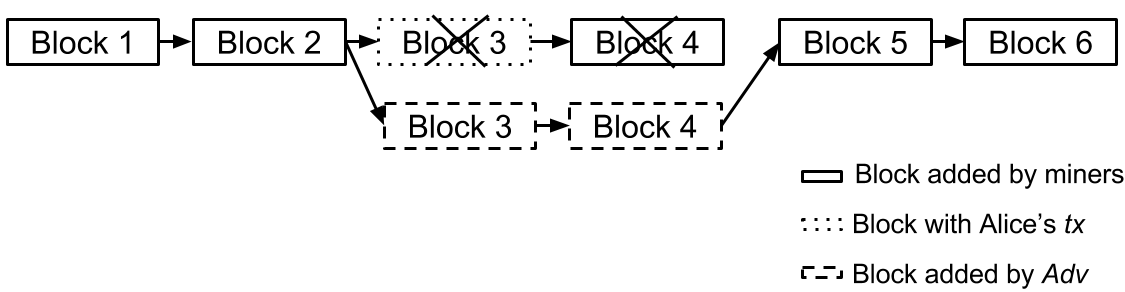}
  \caption{Blacklisting via Punitive Forking}
  \label{fig:ffork}
\end{figure}

\par An adversary with \textit{$>50\%$} hashrate could perform a successful selective blacklisting via \textit{punitive forking}. The objective of punitive forking is to censor the Bitcoin addresses owned by certain people, say \textit{Alice}, and prevent them from spending any of their bitcoins. The strategy to perform the blacklisting (please refer to Figure~\ref{fig:ffork}) is as follows: (i) the adversary with \textit{$>50\%$} network hashrate announces to the Bitcion network that she will not extend on the blockchain containing transactions spending from Alice's Bitcoin address, (ii) if some other miner include a transaction from Alice in a block, the adversary will fork and create a longer proof of work blockchain, (iii) Block containing Alice's transaction now invalidated, and it can never be published, also the miner who added the block with Alice's transaction will lose its block reward. However, a weak adversary that has lower hashrate can still cause delays and inconveniences for Alice's transaction.

\par Punitive forking doesn't work unless you have \textit{$>50\%$} of hashrate. However, there is another strategy to achieve the blacklisting as presented in~\cite{NarayananArvind2016}. In particular, authors present a malicious mining strategy called \textit{feather forking}, in which an attacker announces that she will \textit{attempt} to fork if she sees a block containing Alice's transaction in the blockchain, but she will give up after a while. This is the adversary forks as per its convenience, she will continue to extend its fork until wins (i.e., outraces the main chain), but she gives up (i.e., discard its private fork and continue to extend the main chain) after block with Alice's transaction contains $k$ confirmations. An adversary with total hash power less than 50\% might, with high probability, lose rewards, but it will be able to block the blacklisted transaction with positive probability. Moreover, if the adversary can show that she is determined to block the selected transaction and will perform the retaliatory forking if required, then the rest of the miners will also be motivated to block the blacklisted transactions to avoid the losses, in case, if the attacker retaliates and wins. If this is the case, an attacker might be able to enforce the selective blacklisting with no real cost because other miners are convinced that the attacker will perform a costly feather forking attack if provoked. An attacker performing \textit{feather forking} can also use it to \textit{blackmail} a client by threatening that all her transactions will be put on the blacklist until the client pays the asked ransom coins.

\begin{table*}
\caption {Misbehavior attacks targeting Bitcoin network and entities}
\centering
\scalebox{1.06}{
\begin{tabular}{|L|p{3cm}|p{2cm}|L|p{4cm}|}\hline

\textbf{Attack} & \textbf{Description} & \textbf{Primary targets} & \textbf{Adverse effects} & \textbf{Possible countermeasures} \\ \hline

\textit{Bribery attacks}~\cite{BonneauJ16} & adversary bribe miners to mine on her behalf & miners and merchants & increases probability of a double spend or block withholding & increase the rewards for honest miners, make aware the miners to the long-term losses of bribery~\cite{BonneauJ16}\\\hline

\textit{Refund attacks}~\cite{PatrickMcCorry2016} & adversary exploits the refund policies of existing payment processors & sellers or merchants, users & merchant losses money while honest users might lose their reputation & publicly verifiable evidence~\cite{PatrickMcCorry2016}\\\hline

\textit{Punitive and Feather forking}~\cite{NarayananArvind2016}~\cite{AAMiller2013} & dishonest miners blacklist transactions of specific address & users & freeze the bitcoins of user for forever & remains an open challenge\\\hline

\textit{Transaction malleability}~\cite{AndrychowiczMar2015}~\cite{DeckerChr2014} & adversary change the TXID without invalidating the transaction & Bitcoin exchange centers & exchanges loss funds due to increase in double deposit or double withdrawal instances & multiple metrics for transaction verification~\cite{PPWuille2014}, malleability-resilient ``refund'' transaction~\cite{AndrychowiczMar2015} \\\hline

\textit{Wallet theft}~\cite{wallet2014} & adversary stole or destroy private key of users & individual users or businesses & bitcoins in the wallet are lost & threshold signature based two-factor security~\cite{GennaroRosario2016}~\cite{Goldfederr2014}, hardware wallets~\cite{BamertTobias2014}, TrustZone-backed Bitcoin wallet~\cite{GentilalMiraje2017}, Password-Protected Secret
Sharing (PPSS)~\cite{Jareck2016} \\\hline

\textit{Time jacking}~\cite{corbixgwelt2011} & adversary speed-up the majority of miner's clock & miners & isolate a miner and waste its resources, influence the mining difficulty calculation process & constraint tolerance ranges~\cite{corbixgwelt2011}, network time protocol (NTP) or time sampling on the values received from trusted peers~\cite{DMills2010}\\\hline

\textit{DDoS}~\cite{VasekMarie2014}~\cite{JohnsonBenj2014} & a collaborative attack to exhaust network resources & Bitcoin network, businesses, miners, and users & deny services to honest users/miners, isolate or drive away the miners & Proof-of-Activity (PoA) protocol~\cite{Bentoviddo2014}, fast verification signature based authentication\\\hline

\textit{Sybil}~\cite{Douceur2002} & adversary creates multiple virtual identities & Bitcoin network, miners, users & facilitates time jacking, DDoS, and double spending attacks, threatens user privacy & Xim (a two-party mixing protocol)~\cite{Bissias2014}\\\hline

\textit{Eclipse or netsplit}~\cite{HeilmanEthan2015} & adversary monopolizes all incoming and outgoing connections of victim & miners, users & inconsistent view of the network and blockchain, enable double spends with more than one confirmation & use whitelists, disabling incoming connections~\cite{HeilmanEthan2015}\\\hline

\textit{Tampering}~\cite{GervaisArt2015} & delay the propagation of transactions and blocks to specific nodes & miners, users & mount DoS attacks, wrongfully increase mining advantage, double spend & improve block request management system~\cite{GervaisArt2015}\\\hline

\textit{Routing attacks}~\cite{hijackbtc2017} & isolate a set of nodes from the Bitcoin network, delaying block propagation & miners, users & denial of service attack, increases possibility of 0-confirmation double spends, increases fork rate, waste the mining power of the pools & increase the diversity of node connections, monitor round-trip time, use gateways in different ASes~\cite{hijackbtc2017} \\\hline

\textit{Deanonymization}~\cite{Koshy2014}~\cite{Biryukov2014} & linking IP addresses with a Bitcoin wallet & users & user privacy violation & mixing services~\cite{Danezis2003}, CoinJoin~\cite{GmMaxwell2013}, CoinShuffle~\cite{RuffingTim2014} \\\hline

\end{tabular}}
\label{table:2}
\end{table*}

\subsection{Client-side Security Threats}
\label{sec:Bit_client}

The huge increase in the popularity of bitcoins encouraged a large number of new users to join the network. Each Bitcoin client posses a set of private-public keys in order to access its account or wallet. Hence, it is desirable to have the key management techniques that are secure, yet usable. This is due to the fact that unlike many other applications of cryptography if the keys of a client are lost or compromised, the client will suffer immediate and irrevocable monetary losses. To use the bitcoins, a user needs to install a wallet on her desktop or mobile device. The wallet stores the set of private-public keys associated with the owner of the wallet, thus it is essential to take protective actions to secure the wallet. The \textit{wallet thefts} are mainly performed using mechanisms that include system hacking, installation of buggy software, and incorrect usage of the wallet. 

\par Bitcoin protocol relies heavily on elliptic curve cryptography~\cite{VictorMiller1986} for securing the transactions. In particular, Bitcoin uses elliptic curve digital signature algorithm (ECDSA) which is standardized by NIST~\cite{PGallagher2013} for signing the transactions. For instance, consider the standard ``Pay-to-PubKeyHash'' (P2PKH) transaction script in which the user needs to provide her public key and the signature (using her private key) to prove the ownership. To generate a signature, the user chooses a per-signature random value. For security reason, this value must be kept secret, and it should be different for every other transaction. Repeating per-signature value risks the private key computation, as it has been shown in~\cite{HowgraveGraham2001} that even partially bit-wise equal random values suffice to derive a user's private key. Therefore, it is essential for increasing the security of ECDSA to use highly random and distinct per-signature values for every transaction signature. The inspection of the blockchain for instances, in which the same public key uses the same signature nonces for multiple times has been reported by the authors in~\cite{Bos2014}. In particular, the authors report that there are 158 public keys which have reused the signature nonce in more than one transaction signature, thus making it possible to derive user's private key. Recently, authors in~\cite{Giechaskielll2016} present a systematic analysis of the effects of broken primitives on Bitcoin. Authors highlight the fact that in the current Bitcoin system has no migration plans in-place for both the broken hash and the broken signature scheme, i.e., the Bitcoins RIPEMD160, SHA256, and ECDSA techniques are vulnerable to various security threats such as collision attacks~\cite{Hochj2008}. The authors in~\cite{Giechaskielll2016} found that the main vectors of attack on bitcoins involve collisions on the main hash or attacking the signature scheme, which directly enables coin stealing. However, a break of the address hash has minimal impact, as addresses do not meaningfully protect the privacy of a user. 

\begin{table*}
\large
\caption {Bitcoin wallets}
\centering
\scalebox{0.65}{
\begin{tabular}{|L|p{1.9cm}|p{1.9cm}|p{1.9cm}|p{1.9cm}|p{1.9cm}|p{1.9cm}|p{1.9cm}|p{1.9cm}|p{1.9cm}|p{2.6cm}|}\hline
\textbf{} & \textbf{Coinbase} & \textbf{Blockchain} & \textbf{TREZOR} & \textbf{Exodus} & \textbf{MyCelium} & \textbf{Bitcoin Core} & \textbf{MultiBit HD} & \textbf{Electrum} & \textbf{Copay} & \textbf{Armory}\\ \hline

\textbf{Wallet type} & Hot wallet & Hot wallet  & Hardware wallet  & Hot wallet  & Hot wallet  & Hot wallet  & Hot wallet  & Hot wallet  & Multisig  & Varies \\\hline
\textbf{Web interface} & Yes & Yes & Yes & No & No & No  & No  & No  & Yes  & No  \\\hline
\textbf{Mobile app}  & Yes & Yes & No & No & Yes & No & No  & No  & Yes  & No  \\\hline
\textbf{Desktop client} & No & No & No & Yes & No & Yes & Yes & Yes & Yes  & Yes\\\hline
\textbf{Independent wallet}  & No & No  & Yes & Yes & Yes & Yes & Yes & Yes & Yes  & Yes\\\hline
\textbf{Privacy}  & Moderate & Weak & Variable & Good & Good & Good & Moderate & Good & Good & Good\\\hline
\textbf{Security}  & Good & Good & Good & Good & Good & Good  & Good & Moderate  & Good& Good/Moderate\\\hline

\end{tabular}}
\label{T:Wallet}
\end{table*}

\par Unlike most of the online payment systems that rely on login details consisting of the password and other confidential details for user authentication, Bitcoin relies on public key cryptography. This raises the issues of the secure storage and management of the user keys. Over the years, various type of wallet implementations are researched to obtain secure storage of the user keys, it includes software, online or hosted, hardware or offline, paper and brain wallets. Table~\ref{T:Wallet} shows a number of popular wallets and their main features. Coinbase (coinbase.com), an online wallet is most popular due to its desirable features which it provides to the clients that include: (i) a web interface using which the wallet can be assessed with a browser and Internet connection, (ii) a mobile app that allows access to wallet through mobile devices, (iii) an access to Coinbase do not require a client software and it is independent in nature due to which the wallet providers does not have any control over the funds stored in a client's wallet, and (iv) a moderate level of security and privacy. The \textit{Copay} wallet allows multiple users to be associated with the same wallet, while the \textit{Armory} wallet works in online as well as in offline mode. The wallet providers have to find an adequate trade-off between usability and security while introducing a new wallet into the market. For instance, an online wallet is more susceptible to thefts compared to hardware wallets~\cite{BamertTobias2014} as later are not connected to the Internet, but at the same time hardware wallets lacks usability. If done right, there exist more advanced and secure ways to store the user keys called \textit{paper} and \textit{brain} wallets. As their name indicates, in the paper wallet the keys are written on a document which is stored at some physical location analogizes the cash money storage system, while in brain wallet the keys are stored in the client’s mind in the form of a small passphrase. The passphrase if memorized correctly is then used to generate the correct private key.

\par To avoid the aforementioned risks such as managing cryptographic keys~\cite{Eskandariii2015}, lost or stolen devices, equipment failure, Bitcoin-specific malware~\cite{PLitke2014}, to name a few, that are associated while storing the bitcoins in a wallet, many users might prefer to keep their coins with online exchanges. However, storing the holdings with an exchange makes the users vulnerable to the exchange systems. For instance, one of the most notorious events in the Bitcoin history is the breakdown and ongoing bankruptcy of the oldest and largest exchange called \text{Mt. Gox}, which lost over 450 millions of dollars. Moreover, a few other exchanges have lost their customers’ bitcoins and declared bankruptcy due to external or internal theft, or technical mistakes~\cite{Mooreme2013}. Although, the vulnerability of an exchange system to the disastrous losses can never be fully avoided or mitigated, therefore the authors in~\cite{DagherGa2015} presents \textit{Provisions}, which is a privacy-preserving proof of solvency for Bitcoin exchanges. Provision is a sensible safeguard that requires the periodic demonstrations from the exchanges to show that they control enough bitcoins to settle all of its customers’ accounts.

\subsection{Bitcoin Network Attacks}
\label{sec:Bit_net}

In this section, we will discuss those attacks in the Bitcoin that exploits the existing vulnerabilities in the implementation and design of the Bitcoin protocols and its peer-to-peer communication networking protocols. We will start our discussion with the most common networking attack called \textit{Distributed Denial-of-Service} (DDoS) which targets Bitcoin currency exchanges, mining pools, eWallets, and other financial services in Bitcoin. Due to the distributed nature of Bitcoin network and its consensus protocol, launching a DoS attack has no or minimal adverse effect on network functionalities, hence attackers have to lunch a powerful DDoS to disturb the networking tasks. Unlike DoS attack, in which a single attacker carried out the attack, in DDoS, multiple attackers launch the attack simultaneously. DDoS attacks are inexpensive to carry out, yet quite disruptive in nature. Malicious miners can perform a DDoS (by having access to a distributed Botnet) on competing miners, effectively taking the competing miners out of the network and increasing the malicious miner’s effective hashrate. In these attacks, the adversary exhausts the network resources in order to disrupt their access to genuine users. For example, an honest miner is congested with the requests (such as fake transactions) from a large number of clients acting under the control of an adversary. After a while, the miner will likely to start discarding all the incoming inputs/requests including requests from honest clients. In~\cite{VasekMarie2014}, authors provide a comprehensive empirical analysis of DDoS attacks in the Bitcoin by documenting the following main facts: 142 unique DDoS attacks on 40 Bitcoin services and 7\% of all known operators were victims of these attacks. The paper also states that the majority of DDoS attack targets the exchange services and large mining pools because a successful attack on these will earn huge revenue for the adversary as compared to attacking an individual or small mining pools. 

\par In~\cite{JohnsonBenj2014}, authors explore the trade-off between the two mining pool related strategies using a series of game-theoretical models. The first strategy called \textit{construction}, in which a mining pool invests in increasing its mining capacity in order to increase the likelihood of winning the next race. While in the second strategy called \textit{destruction}, in which the mining pool launches a costly DDoS attack to lower the expected success rate of the competing mining pools. The majority of the DDoS attacks target large organizations due to bulk ransom motives. Companies like CoinWallet and BitQuick were forced to shut down only after few months of their launch due to the effects of continuous DDoS attacks. As stated above that DDoS attack take various forms, one of which is to discourage a miner so that it will withdraw itself from the mining process. For instance, an attacker displays to a colleague miner that it is more powerful, and it can snatch the reward of mining, and it is the obvious winner of the mining process. An honest miner backoffs since its chances of winning is less. In this way, an adversary will be successful in removing individual miners as well as small pools from the mining network, thus imposing a DDoS attack on the network~\cite{JohnsonBenj2014}. Moreover, in~\cite{eudecker2015}, authors propose network partitioning in Bitcoin, hence isolating the honest nodes from the network by reducing their reputation.

\par Now we discuss the so-called \textit{Malleability attacks}~\cite{DeckerChr2014}, which also facilitates the DDoS attacks in Bitcoin. For instance, by using a \textit{Malleability attack} an adversary clogs the transaction queue~\cite{malla2014}. This queue consists of all the pending transactions which are about to be serviced in the network. Meanwhile, an adversary puts in bogus transactions with the high priority depicting itself to be highest incentive payer for the miners. When the miners try to verify these transactions, they will find that these are the false transaction, and but by this time they have already spent a considerable amount of time in verifying these false transactions. This attack wastes the time and resources of the miners and the network~\cite{malla2017}. Malleability is defined in terms of cryptography by~\cite{DeckerChr2014}. A cryptographic primitive is considered malleable, if its output $Y$ can be ``mauled'' to some ``similar'' value $Y'$ by an adversary who is unaware of the cryptographic secrets that were used to develop $Y$. 

\par In~\cite{AndrychowiczMar2015}, another form of \textit{malleability} attack called \emph{transaction malleability} is introduced. Suppose that a transaction $T_{A \rightarrow B}^n$ which transfers $n$ bitcoins from $A's$ wallet to $B's$ wallet. With \textit{transaction malleability} it is possible to create another $T'$ that is syntactically different (i.e., $T_{A \rightarrow B}^n$ and $T'$ has different transaction hash ID $T_x^{id}$) from $T_{A \rightarrow B}^n$, although semantically it is identical (i.e. $T'$ also transfers $n$ coins from wallet $A$ to $B$). An adversary can perform \textit{transaction malleability} without even knowing the private key of $A$. On a high level, \textit{transaction malleability} refers to a bug in the original Bitcoin protocol which allows the aforementioned behavior in the network possible. The main reason of the success of this attack is that, in Bitcoin each transaction is uniquely identified by its $T_x^{id}$, hence in some cases $T'$ will be considered a different transaction than $T_{A \rightarrow B}^n$. 
 
\par In Bitcoin, certainly, the transaction malleability is not desirable, but it does not cause any damage to the system until an adversary exploits its behavior and make someone believe that a transaction has been failed. However, after a while, the same transaction gets published in the global blockchain. This might lead to a possible double spend, but it is particularly more relevant while targeting the Bitcoin exchanges which holds a significant amount of coins. This is because it allows the users to buy and sell bitcoins in exchange for cash money or altcoins. The Bitcoin’s reference implementation is immune to the transaction malleability because it uses previous transaction's outputs as an indication for the successfully issued transactions. However, few exchanges use a custom implementation and were apparently vulnerable. For instance, Mt. Gox (a popular exchange) issued a statement in the early days of Bitcoin that they were attacked due to transaction malleability, therefore they are forced to halt withdrawals and freezing clients account. The attack that MtGox claimed to be the victim proceeds as follows: (i) an dishonest client $C_d$ deposits $n$ coins in his MtGox account, (ii) $C_d$ sends a transaction $T$ to MtGox asking to transfer her $n$ coins back, (iii) MtGox issues a transaction $T'$ which transfers $n$ coins to $C_d$, (iv) $C_d$ performs the malleability attack, obtaining $T'$ that is semantically equivalent to $T$ but has a different $T_x^{id}$, now assume that $T'$ gets included into the blockchain instead of $T$, (v) $C_d$ complains to MtGox that the transaction $T$ was not successful, (vi) MtGox performs an internal check, and it will not found a successful transaction with the $T_x^{id}$, thus MtGox credits the money back to the user's wallet. Hence effectively $C_d$ is able to withdraw her coins twice. The whole problem is in the above Step (vi), where MtGox should have searched not for the transaction with $T_x^{id}$ of $T$, but for any transaction semantically equivalent to $T$.

\par For the first time, authors in~\cite{hijackbtc2017} present the impact of routing attacks on Bitcoin network by considering both small and large scale attacks. The paper shows that two key properties of Bitcoin networks which includes, the ease of routing manipulation, and the rapidly increasing centralization of Bitcoin in terms of mining power and routing, makes the routing attacks practical. More specifically, the key observations suggest that any adversary with few ($<100$) hijacked BGP prefixes could partition nearly 50\% of the mining power, even when considering that mining pools are heavily multi-homed. The research also shows that attackers on acting as intermediate nodes can considerably slow down block propagation by interfering with few key Bitcoin messages. Authors back their claims by demonstrating the feasibility of each attack against the deployed Bitcoin software, and quantify their effect on the current Bitcoin topology using data collected from a Bitcoin supernode combined with BGP routing data. Furthermore, to prevent the effect of aforementioned attacks in practice,  both short and long-term countermeasures, some of which can be deployed immediately are suggested.

\par Due to the vulnerabilities that exist in the refund policies of the current Bitcoin payment protocol, a malicious user can perform the so-called \textit{Refund attacks}. In~\cite{PatrickMcCorry2016}, authors present the successful implementation of the refund attacks on $BIP70$ payment protocol. BIP70 is a Bitcoin community-accepted standard payment protocol that governs how vendors and customers perform payments in Bitcoin. Most of the major wallets use BIP70 for bitcoins exchange, and the two dominant Payment Processors called \textit{Coinbase} and \textit{BitPay}, who uses BIP70 and collectively they provide the infrastructure for accepting bitcoins as a form of payment to more than 100,000 vendors. The authors propose two types of refund attacks called \textit{Silkroad Trader attack} which highlights an authentication vulnerability in the BIP70, and \textit{Marketplace Trader attack} which exploits the refund policies of existing payment processors. The brief description of both these refund attacks is as follows.
\begin{itemize}
\item In \textit{Silkroad attack}, a customer is under the control of an ill trader. When a customer starts trading with the merchant its address is revealed to the ill trader. When the transaction is finished, the adversary initiates the attack by inserting the customers' address as the refund address and send a refund request to the merchant. The merchant sends the amount to the ill merchant, thus gets cheated without receiving a refund from the other side. During this whole process of refund between the merchant and the ill trader, the customer is not at all aware of the fraud that is happening in her name.
\item The \textit{Marketplace trader attack} is a typical case of the man-in-the-middle attack. In this, the adversary setup an attractive webpage where she attracts the customer who falls victim in the later stages. The attacker depicts herself as a trusted party by making payments through trust-able merchants like CeX. When a customer clicks the webpage, accidentally reveals her address among the other identities that are sufficient to perform malpractice by the rogue trader with the false webpage. When customer purchase products, a payment page is sent which is a legitimate payment exchange merchant. The end merchant is connected to the adversary's webpage and meanwhile, the details of the customer would have been already revealed to the attacker through an external email communication according to the Bitcoin refund policies. After the transaction, the middle adversary claims a refund on behalf of the customer and the refund amount will be sent to the rogue adversary's account. Hence, the legitimate customer will not be aware of the fraud process but the merchant loses his bitcoins~\cite{PatrickMcCorry2016}.
\end{itemize}
Later, both these attacks have been acknowledged by Coinbase and Bitpay with temporary mitigation measures put in place. However, the authors claim that to fully address the identified issues will require revising the BIP70 standard. 

\par Yet another attack on the Bitcoin networks is called \textit{Time jacking attack}~\cite{corbixgwelt2011}. In Bitcoin network, all the participating nodes internally maintain a time counter that represents the network time. The value of the time counter is based on the median time of a node's peers, and it is sent in the version message when peers first connect. However, if the median time differs by more than 70 minutes from the system time, the network time counter reverts to the system time. An adversary could plant multiple fake peers in the network and all these peers will report inaccurate timestamps, hence it can potentially slow down or speed up a node's network time counter. An advanced form of this attack would involve speeding up the clocks of a majority of the miners while slowing down the target's clock. Since the time value can be skewed by at most 70 minutes, the difference between the nodes time would be 140 minutes~\cite{corbixgwelt2011}. Furthermore, by announcing inaccurate timestamps, an attacker can alter a node's network time counter and deceive it into accepting an alternate blockchain because the creation of new blocks heavily depends on network time counters. This attack significantly increases the possibility of the following misbehaviors: a successful double spending attack, exhaust computational resources of miners, and slow down the transaction confirmation rate. 

\begin{figure}[h]
\centering
\includegraphics[scale = 0.45]{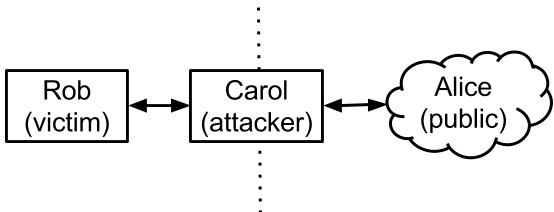}
  \caption{Eclipse attack}
  \label{fig:eclipse}
\end{figure}

\par Apart from the aforementioned major attacks on Bitcoin protocol and network, there are few other minor attacks that we have summarized below. 
\begin{itemize}
\item \textit{Sybil Attack:} A type of attack where attacker installs dummy helper nodes and tries to compromise a part of the Bitcoin network. A sybil attack~\cite{Douceur2002} is a collaborative attack performed by a group of compromised nodes. Also, an attacker may change its identity and may launch a collusion attack with the helper nodes. An attacker tries to isolate the user and disconnect the transactions initiated by the user or a user will be made to choose only those blocks that are governed by the attacker. If no nodes in the network confirm a transaction that input can be used for double spending attack. An intruder with her helper nodes can perform a collaborated timing attack, hence it can hamper a low latency encryption associated with the network. The other version of this attack where the attacker tries to track back the nodes and wallets involved in the transaction is discussed in~\cite{Bissias2014}.

\item \textit{Eclipse attack:} In this attack~\cite{HeilmanEthan2015}, an adversary manipulates a victim peer, and it force network partition (as shown in Figure~\ref{fig:eclipse}) between the public network and a specific miner (victim). The IP addresses to which the victim user connects are blocked or diverted towards an adversary~\cite{HeilmanEthan2015}. In addition, an attacker can hold multiple IP addresses to spoof the victims from the network. An attacker may deploy helpers and launch other attacks on the network such as $N-confirmation$ double spending and selfish mining. The attack could be of two type: (i) Infrastructure attacks, where attack is on the ISP (Internet Service Provider) which holds numerous contiguous addresses, hence it can manipulate multiple addresses that connect peer-to-peer in the network, and (ii) botnet attacks, where an adversary can manipulate addresses in a particular range, especially in small companies which own their private set of IP addresses. In both the cases, an adversary can manipulate the peers in the Bitcoin network.

\item \textit{Tampering:} In a Bitcoin network, after mining a block the miners broadcast the information about newly mined blocks. New transactions will be broadcast from time to time in the network. The network assumes that the messages will reach to the other nodes in the network with a good speed. However, authors in~\cite{GervaisArt2015} ground this assumption and proved that the adversary can induce delays in the broadcast packets by introducing congestion in the network or making a victim node busy by sending requests to all its ports. Such type of tampering can become a root cause for other types of attacks in the network.
\end{itemize}

\subsection{Practical attack incidents on Bitcoin}
\label{sec:real-world}
In this section, we briefly present the existing real-world security breaches/incidents that have affected adversely to Bitcoin and its associated technologies, such as blockchain and PoW based consensus protocol. From the start, bitcoin fans occasionally mentioned about different security, typically discussing things like the 51\% attack, quantum computer strikes, or an extreme denial of service onslaught from some central bank or government entity. However, these days the word \textit{attack} is used a bit more loosely than ever, as the scaling debate has made people believe almost everything is a Bitcoin network invasion.

\par One of the biggest attacks in the history of Bitcoin have targeted \textit{Mt. Gox}, the largest Bitcoin exchange, in which a year's long hacking effort to get into Mt. Gox culminated in the loss of 744,408 bitcoins. However, the legitimacy of attack was not completely confirmed, but it was enough to make Mt. Gox to shut down and the value of bitcoins to slide to a three-month low. In 2013, another attack called \textit{Silk Road}, the worlds largest online anonymous market famous for its wide collection of illicit drugs and its use of Tor and Bitcoin to protect its user's privacy, reports that it is currently being subjected to what may be the most powerful distributed denial-of-service attack against the site to date. In the official statement from the company the following was stated, ``The initial investigations indicate that a vendor exploited a recently discovered vulnerability in the Bitcoin protocol known as \textit{transaction malleability} to repeatedly withdraw coins from our system until it was completely empty''. Although transaction malleability is now being addressed by \textit{segwit}, the loss it caused was far too small with the main issue seemingly being at a human level, rather than protocol level. In the same year, \textit{Sheep Marketplace}, one of the leading anonymous websites also announces that they have been hacked by an anonymous vendor \textit{EBOOK101} who stole 5400 bitcoins. However, in all the aforementioned, it remains unclear that whether there is any hacked happened or it is just a fraud by the owners to stole the bitcoins.

\par \textit{Bitstamp}, an alternative to MT Gox, increasing its market-share while Gox went under were hacked out of around 5 million dollars in 2015. The theft seems to have been a sophisticated attack, with phishing emails targeting bitstamp’s personnel. However, as the theft was limited to just hot wallets, they were able to fully cover it, leading to no direct customer losses. \textit{Poloniex} is one of the biggest altcoin exchange with trading volumes of 100,000 BTC or more per day, lost their 12.3\% of bitcoins in March 2014. The hack was executed by just clicking withdrawal more than once. As it can be concluded from the above discussion that the attackers always target the popular exchanges to increase their profit. However, it does not implies that individual users are not targeted, it's just that the small attacks go unnoticed. Recently, in August 2016, \textit{BitFinex}, which a popular cryptocurrency exchange suffered a hack due to their wallet vulnerability, and as a result around 120000 bitcoins were stolen.

\par From the nature of the aforementioned attacks, it can be concluded that security is a vital concern and biggest weakness for cryptocurrency marketplaces and exchanges. In particular, as the number of bitcoins stored and their value has skyrocketed over the last year, bitcoins digital wallets have increasingly become a target for hackers. At the social level, what is obvious and does not need mentioning (although some, amazingly, dispute it) is that individuals who handle our bitcoins should be public figures with their full background on display for otherwise they cannot be held accountable. Lacking such accountability, hundreds of millions, understandably, is far too tempting as we have often seen. An equally important point is that bitcoin security is very hard. Exchanges, in particular, require highly experienced developers who are fully familiar with the bitcoin protocol, the many aspects of exchange coding and how to secure hard digital assets for, to truly secure bitcoin, exchanges need layers and layers amounting to metaphorical armed guards defending iron gates with vaults deep underground behind a thousand doors.


\section{Security: Countermeasures for Bitcoin Attacks}
\label{sec:Bit_defense}

In this section, we discuss the state-of-the-art security solutions that provides possible countermeasures for the array of attacks (please refer to Section~\ref{sec:Bit_security}) on Bitcoin and its underlying technologies.

\subsection{No more double spending}
\label{sec:Bit_defense_DB}

The transaction propagation and mining processes in Bitcoin provide an inherently high level of protection against double spending. This is achieved by enforcing a simple rule that only unspent outputs from the previous transaction may be used in the input of a next transaction, and the order of transactions is specified by their chronological order in the blockchain which is enforced using strong cryptography techniques. This boils down to a distributed consensus algorithm and time-stamping. In particular, the default solution that provides resistance to double spending in Bitcoin is its use of Proof-of-work (PoW) based consensus algorithm, which limits the capabilities of an adversary in terms of, the computational resources available to an adversary and the percentage of honest miners in the network. More specifically, the purpose of the PoW is to reach consensus in the network regarding the blockchain history, thereby synchronizing the transactions or blocks and making the users secure against double-spending attacks. Moreover, the concept of PoW protect the network against being vulnerable to sybil attack because a successful sybil attack could sabotage the functionality of consensus algorithm and leads to possible double spending attack.

\par In general, double spending could be dealt in two possibles ways: (i) detect a double spending instance by monitoring the blockchain progress, and once detected, identify the adversary and take adequate actions, or (ii) use preventive measures. The former approach works well in the traditional centralized online banking system, but in Bitcoin, it's not suitable due to the use of continuously varying Bitcoin address, thus it provides anonymity to users, and the lack of transaction rollback scheme once it is successfully added in the blockchain. Therefore, the latter approach, i.e., prevent double spend is desirable in Bitcoin. 

\par The most effective yet simple way to prevent a double spend is to wait for a multiple numbers of confirmations before delivering goods or services to the payee. In particular, the possibility of a successful double spend decreases with increase in the number of confirmations received. Of course, the longer back transactions lie in the blockchain, the more blocks need to be caught up until a malicious chain gets accepted in the network. This limits attacker from possible revise the history of transactions in the chain. For instance, unconfirmed bitcoin transaction (zero block transaction) has a high risk of double spend, while a transaction with atleast one confirmation has \textit{statically} zero risks of double spend, and a transaction with six confirmations are commonly considered steady, hence has zero risks of double spend. In Bitcoin, the classic bitcoin client will show a transaction as \textit{not unconfirmed} until the transaction is six blocks deep\footnote{Each new block that will be put on top of a block containing the desired transaction will result in the generation of a confirmation for the desired transaction.} in the blockchain. However, waiting of six transactions (about one hour) might not be suitable for various applications such as fast payment systems, e.g., Alice is very hungry and she wants to buy a snack with bitcoins. There is nothing special about the choice of the default \textit{safe} confirmation value, i.e., six confirmations. Its choice is based on the assumption that an adversary is unlikely to control more than 10\% of the mining power, and that a negligible risk lower than 0.1\% is acceptable. This means that on one hand, the six confirmations are overkill for casual attackers, while at the same time it is powerless against more dedicated attackers with much more than 10\% mining power.  

\par Authors in~\cite{KarameGhassan2012} evaluate three techniques that can be used to detect a possible double spending in fast payment systems. The three techniques are as follow: \textit{listening period}, \textit{inserting observers}, and \textit{forwarding double spending attempts}. In the first technique, the vendor associates a listening period with each received transaction, and it monitors all the receiving transactions during this period. The vendor only delivers the product, if it does not see any attempt of double spending during its listening period. The \textit{inserting observers} technique naturally extends the first technique based on the adoption of a listening period would be for the vendor to insert a set of nodes (i.e., ``observers'') under its control within the Bitcoin network. These observers will directly relay all the transactions to the vendor that they receive from the network. In this way, with the help of the observers, the vendor is able to \textit{see} more number of transactions in the network during its \textit{listening period}, thus increases the chances of detecting a double spend. The third technique (i.e., \textit{forwarding double spending attempts}) requires each Bitcoin peer to forward all transactions that attempt to double spend instead of discarding them so that the vendor can receive such a transactions on time (i.e., before releasing the product). With this approach, whenever a peer receives a new transaction, it checks whether the transaction is an attempt to double spend, if so then peer forward the transaction to their neighbors (without adding it to their memory pools).

\par Recently, the hash power of a pool called \textit{GHash.IO} reached $54\%$ for a day (i.e., it exceeds the theoretical attack threshold of $51\%$). Although the \textit{GHash.IO} remained honest by transferring a part of its mining power to other pools. However, the incentives that motivate an adversary to create large pools remains in the network, always looking for a chance to wrongful gain and disrupt the network. Therefore, a method to prevent the formation of large pools called \textit{Two phase Proof-of-Work} (2P-PoW) has been proposed in~\cite{MartijnBastiaan2015}. The authors propose a second proof-of-work (say $Y$) on top of the traditional proof-of-work (say $X$) of the block header. $Y$ signs the produced header with the private key controlling the payout address. Similar to existing hashing procedures this signature must meet a target set by the network, hence the use of $Y$ forces pool managers to distribute their private key to their clients if the manager wants to retain the same level of decentralization. However, if a manager would naively share its private key, all clients would be authorized to move funds from the payout address to any destination. Pool managers unwilling to share their private key needs to install mining equipment required to solve $Y$ in a timely manner. It is estimated that GHash.IO owns only a small percentage of the network's computing power in terms of hardware, as the pool shrank significantly after public outrage. Depending on the difficulty of $Y's$ cryptographic puzzle this would only allow a certain number of untrusted individuals to join. In this way, as GHash.IO is a public pool, severely limit its size.

\par Authors in~\cite{Ruffingr2015} propose the use of decentralized non-equivocation contracts, to detect the double spending and penalize the malicious payer. The basic idea of non-equivocation contracts is that the payer locks some bitcoins in a deposit when he initiates a transaction with the payee. If the payer double spends, a cryptographic primitive called accountable assertions can be used to reveal his Bitcoin credentials for the deposit. Thus, the malicious payer could be penalized by the loss of deposit coins. However, such decentralized non-equivocation contracts are subjected to collusion attacks where the payer colludes with the beneficiary of the deposit and transfers the Bitcoin deposit back to himself when he double spends, resulting in no penalties. On the other hand, even if the beneficiary behaves honestly, the victim payee cannot get any compensation directly from the deposit in the original design. To prevent such collusion attacks, authors in~\cite{XYu2017} design fair deposits for Bitcoin transactions to defend against double-spending. The fair deposits ensure that the payer will be penalized by the loss of his deposit coins if he double spends and the victim payee’s loss will be compensated. The proposed protocol uses the assertion scheme from~\cite{Ruffingr2015}. In particular, the beneficiary can recover the payer’s secret key if the payer double spends. However, to ensure that the payee’s loss can be compensated if the payer double spends, in addition to a signature generated with the payer’s secret key, a signature generated with the payee’s secret key is required for the release of the compensation locked in the deposit. Meanwhile, the incentive for the beneficiary is also guaranteed in the deposit.

\par Another solution to control double spending was proposed in~\cite{Karamea2012} where all the participating users deposit a safety amount similar to an agreement. If an attacker tries to double spend and it is detected, the deposit amount will be deducted and it is given to the victim who encountered the loss. Due to the punishing attribute of the network, the attack can be controlled. In~\cite{LearBahack2013}, authors suggest a countermeasure by prohibiting the merchant to accept incoming connections, thus an adversary cannot directly send a transaction to the merchant. This forces the adversary to broadcast the transaction over the Bitcoin network, and this ensures that the transaction will end up in the local view of all the miners that forwards it. Later if the adversary tries to double spend the miners will know about it and take primitive actions in future.

\par Solution for $50\%$ attack is presented in~\cite{LearBahack2013}. The authors provide countermeasures for two variants of $50\%$ attack namely: \textit{block discarding attack} and \textit{difficulty rising attack}. In block discarding attack, an adversary has control over a set of nodes in the network, called \textit{supporters}. The adversary and her supporters purposefully add a delay in the propagation of the legitimately discovered blocks, and the attacker advertises her block selfishly. Hence, the advertiser's blockchain will increase, and the other blocks due to delay get less attention. The delay becomes worse as the number of supporter increases. The solution for this attack is fixing the punishment for the advertisers or the misbehaving miners. Every node is asked to pay a deposit amount, and the nodes who misbehave are punished by dissolving the deposit amount of the concerned. This amount is distributed among the nodes who informs about the misbehaving node in the network. While in difficulty rising attack, the attacker manipulates the network and slowly raises the difficulty level for the miners. An attacker poses a threat to the network by controlling high hash-power compared with other nodes in the network. The solution to this attack is same as that of block discarding attack. In~\cite{sybil51}, authors propose a method called ``proof-of-reputation'', where the honest miners will get a token based on the current market value. The number of tokens issued can vary with the market value. If the miner has the token, he will be reputed in the mining market pool. The token has a value, and according to which the coins are deposited from all the miners from time to time and is fixed by the network. More the reputation of the miner's chain, more the other blocks merge with that chain.

\par For now, it is safe to conclude that there is no solution available in the literature that guarantees the complete protection from double spending in Bitcoin. The existing solutions only make the attack more difficult for adversaries. In particular, double spending is an attack that is well discussed in the Bitcoin community, but very few solutions exist so far, and it remains an open challenge for the researchers. The easiest, yet most powerful way for a vendor to avoid a double spend is to wait for more number of confirmations before accepting a transaction. Therefore, each vendor or merchant of the deals in bitcoins has to set a trade-off between the risk and the product delivery time caused while waiting for an appropriate number of confirmations. Similar to the honest Bitcoin users, there is also a trade-off for the adversary as she needs to consider the expenses (i.e., the loss of computing resources and rewards for the pre-mined blocks) if the attack fails.  

\subsection{Countermeasures for Private Forking and Pool Attacks} 
\label{sec:Bit_defense_MPool}

When a dishonest miner intentionally forks the blockchain by privately mining a set of blocks, it makes the Bitcoin network vulnerable to a wide range of attacks such as selfish mining, block-discarding attack, block withholding attack, bribery attack, to name a few. The aim of these attacks is to cheat the mining incentive system of Bitcoin. Therefore, at any point in time, detecting and mitigating the faulty forks from the set of available forks poses a major challenge for Bitcoin protocol developers. The simplest solution to handle the selfish mining is suggested in~\cite{IttayEyalS13}. The authors propose a simple, backward-compatible change to the Bitcoin protocol. In particular, when a miner encounters the presence of multiple forks of the same length, it will forward this information to all its peers, and it randomly chooses one fork to extend. Hence, each miner implementing the above approach by selecting a random fork to extend. This approach will decrease the selfish pool's ability to increase the probability that other miners will extend their fork. 

\par To further extend the countermeasure presented in~\cite{IttayEyalS13}, authors in~\cite{HeilmanEthan2014} introduce the concept of \textit{Freshness Preferred} (FP), which places the unforgeable timestamps in blocks and prefer blocks with recent timestamps. This approach uses Random Beacons~\cite{RABIN1983} in order to stop miners from using timestamps from the future. As the selfish mining uses strategic block withholding technique, the proposed strategy will decrease the incentives for selfish mining because withheld blocks will lose block races against newly minted or \textit{fresh} blocks. A similar but a more robust solution for selfish mining that requires no changes in existing Bitcoin protocol is proposed in~\cite{ZhangRen2017}. The authors suggest a fork-resolving policy that selectively neglects blocks that are not published in time, and it appreciates blocks that include a pointer to competing blocks of their predecessors. Therefore, if the secretly mined block is not published in the network until a competing block is published, it will contribute to neither or both branches, hence it will not get benefits in winning the fork race. Authors in~\cite{Ren2015} proposes another defense against selfish mining, in which miners need to publish intermediate blocks (or in-blocks). These blocks, although are valid with lower puzzle difficulty, but confer no mining reward onto the miner who discovers one. When a fork happens, miners adopt the branch with the largest total amount of work, rather than the longest chain.   

\par Unlike most of the aforementioned solutions against malicious forking, authors in~\cite{SiamakSolatP16} propose a timestamp-free prevention of block withholding attack called \textit{ZeroBlock}. In ZeroBlock, if a selfish miner keeps a mined block private more than a specified interval called \textit{mat}, than later when this block is published on the network, it will be rejected by the honest miners. The key idea is that each consecutive block must be published in the network, and it should be received by honest miners within a predefined maximum acceptable time for receiving a new block (i.e., \textit{mat} interval). In particular, an honest miner either receives or publishes the next block in the network within the \textit{mat} interval. Otherwise, to prevent the block withholding, the miner itself generates a specific dummy block called \textit{Zeroblock}. These signed dummy Zeroblocks will accompany the solved blocks to prove, that the block is witnessed by the network and that a competing block is absent before miners are able to work on it. For forking attacks that are internal to a pool, authors in~\cite{NicolasT2014} suggest that the only viable option to countermeasure a block withholding attack launched within a pool is that the pool managers should involve \textit{ONLY} miners which are personally known to them, hence they can be trusted. The pool manager should simply dissolve and close a pool as soon as the earning of the pool goes lower than expected from its computational effort.  

\par In~\cite{BonneauJ16}, bribery attack is discussed along with its countermeasure. In bribery, an attacker bribe a miner to rent her computing resources, thus it increases the attackers hash power that it could use to launch various attacks in the network. As a countermeasure, authors suggest the use of anti-payment (i.e, counter-bribing) to pool miners which have value more than what attackers are paying to these miners to perform a malicious behavior. However, the drawback is that a legitimate pool manager has to spend a lot to take miners toward the normal mining routine. In addition, as the number of bribing node or a node's bribe amount increases, the capital requirements for the manager also increases, and as the crypt math becomes more and more difficult the bribe amount increases, hence makes it difficult for the manager to keep the process of counter-bribing active for longer periods.

\subsection{Securing Bitcoin wallets}
\label{sec:Bit_defense_client}

A wallet contains private keys, one for each account~\cite{Eskandariii2015}. These private keys are encrypted using the master key which is a random key, and it is encrypted using AES-256-CBC with a key derived from a passphrase using SHA-512 and OpenSSLs EVP\_BytesToKey~\cite{StevenGoldfederr2016}. Private key combined with the public key generates a digital signature which is used to transact from peer-to-peer. Bitcoin uses ECDSA (Elliptic Curve Digital Signature Algorithm) algorithm for encryption, and it is modified in~\cite{Bos2014} for secret sharing and threshold cryptography.

\par A manual method of wallet protection was proposed by~\cite{coldwelt2016} called ``cold wallet''. A cold wallet is another account that holds the excess of an amount by the user. This method uses two computers (the second computer has to be disconnected from the Internet) and using the Bitcoin wallet software a new private key is generated. The excess amount is sent to this new wallet using the private key of a user. Authors in~\cite{coldwelt2016} claim that if the computer is not connected to the Internet, the hackers will not get to know the keys, hence the wallet safety can be achieved. Securing wallets with new cryptographic algorithms apart from ECDSA is still an open issue and a challenge. In~\cite{tagkeyy2015}, an article states that US government have launched their own Bitcoin networks with multi-factor security which incorporates fingerprint biometrics for wallet protection. A device is a standalone tool same as the size of a credit card. In~\cite{BamertTobias2014}, authors propose \textit{BlueWallet}, a proof-of-concept based hardware token for the authorization of transactions in order to protect the private keys. The concept is similar to the use of the ``cold wallet'', that is, it uses a dedicated hardware not connected to the Internet to store the private keys. The hardware token communicates with the computer (or any other device) that creates the transaction using Bluetooth Low Energy (BLE) and it can review the transaction before signing it. The securely stored private key never leaves the BlueWallet and is only unlocked if the user correctly enters her PIN. BlueWallet provides the desired security at the expense of the usability, as the users have to invest and keep an additional device while making a transaction. 

\par Bitcoin already has a built-in function to increase the security of its wallets called ``multi-signature'', which tightens the security by employing the splitting control technique. For instance, \textit{BitGo}~-~an online wallet which provides 2-of-3 multi-signature transactions to its clients. However, the drawback of using the multi-signature transactions is that it greatly compromises the privacy and anonymity of the user. Authors in~\cite{GennaroRosario2016} propose an efficient and optimal threshold Digital Signature Algorithm (DSA) scheme for securing private keys. The main idea behind the use of threshold signatures proposed in~\cite{GennaroRosario2016} is derived from secret sharing~\cite{Shami1979}, in which the private key is split into shares. Any subset of the shares that is equal to or greater than a predefined threshold is able to reconstruct the private key, but any subset that is smaller will gain no information about the key. The main property of threshold signatures~\cite{Goldfederr2014} is that the key is never revealed because the participants directly construct a signature. Recently, authors in~\cite{GentilalMiraje2017} present a TrustZone\footnote{TrustZone is a technology that is used as an extension of processors and system architectures to increase their security.} based Bitcoin wallet and shows that it is more resilient to the dictionary and side-channel attacks. Although the use of TrustZone makes use of the encrypted storage, hence the writing and reading operations become slower.


\subsection{Securing Bitcoin Networks}
\label{sec:Bit_defense_net}

In this section, we will discuss various existing countermeasures proposed for securing the Bitcoin's core protocols and its peer-to-peer networking infrastructure functionalities against an array of security threats some of which we have discussed in Section~\ref{sec:Bit_net}.

\subsubsection{DDoS Attacks}
In~\cite{JohnsonBenj2014}, authors propose a game theoretic approach for analyzing the DDoS attacks. The game assumes that the pools are in competition with each other because the larger pools are always weighted more than the smaller pools. The game exists between the pools, and each pool tries to increase their computational cost over others, and then it imposes a DDoS attack on the other pools. In this way, authors draw an equilibrium condition between the players and concludes that the larger pools will have more incentives against the smaller pools. In~\cite{IttayEyal14}, authors propose a ``miner's dilemma'', again a game theoretical approach to model the behavior of miners similar to repetitive prisoner's dilemma. There exist a game between the pools. The longest chain dominates over the smaller chains and grabs the rewards by behaving selfishly in the network. Game theory concludes that by performing attacks, the pools actually lose the bitcoins that they are supposed to get when compared it with the case without attacking each other. In particular, this kind of game theory problems is called ``Tragedy of Commons'', where the peers turn out to be rational, selfish and harm other peers for their benefits.

\par In~\cite{Bentoviddo2014}, authors propose Proof-of-Activity (PoA) protocol, which is robust against a DDoS attack that could be launched by broadcasting a large number of invalid blocks in the network. In PoA, each block header is stored with a crypt value and the user that stores the first transaction places this value. These users are called ``stakeholders'' in the network and they are assumed, to be honest. Any subsequent storage of transactions in this block is done if there are valid stakeholders associated with the block. Storage of crypt value is random and more transactions are stored, only if more stake users are associated with the chain. If the length of the chain is more, trustworthiness among other peers increases and more miners get attracted towards the chain. Hence, an adversary cannot place a malicious block or transaction since all the nodes in the network are governed by stakeholders.

\par One possible way to mitigate DDoS attacks is to use the technique discussed in~\cite{cCameloMK14}, which suggests the continuous monitoring of network traffic by using browsers like Tor or any user-defined web service. Applying machine-learning techniques like SVM and clustering will identify which part of the network is behaving ill. Hence that part can be isolated from the network until debugged. Other possible methods to protect against DDoS attacks include: (i) configure the network in a way that malicious packets and requests from unnecessary ports will be prohibited, (ii) implement a third party DoS protection scheme which carefully monitors the network and identify variations in the pattern. We believe that similar approaches could also be implemented in future in Bitcoin networks to countermeasure DoS attacks. 

\subsubsection{Time Jacking and Eclipse Attack}
In this attack an adversary alters the node time, therefore the dependency of a node on network time can be replaced by a hardware oriented system time. The accept time window for transactions at a node has to be reduced, making the node recover quicker from the attacks. \textit{Time jacking} is a dreaded attack that might split the network into multiple parts, hence it can isolate the victim node. A set of techniques is suggested in~\cite{corbixgwelt2011} to avoid time jacking that includes, use of the system time instead of network time to determine the upper limit of block timestamps, tighten the acceptable time ranges, and use only trusted peers. Even a node can be designed to hold multiple timestamps assuming that the attacker may not alter all the timestamps. Furthermore, node timestamps can be made dependent on the blockchain timestamps~\cite{corbixgwelt2011}.

\par In~\cite{HeilmanEthan2015}, authors provide techniques to combat eclipse attack which uses an additional procedure to store the IP addresses that are trustworthy. If the users are connected to other peers in the network, these peers are stored in ``tried'' variable. The connection of the user with the peers is dependent on the threshold of the trust factor, which varies from time to time. The users can have special intrusion detection system to check the misbehaving nodes in the network. The addresses which misbehave in the network could be banned from connections. These features can prevent the users from an eclipse attack. In particular, having a check on the incoming and outgoing connections from the node can reduce the effect of an eclipse attack.

\subsubsection{Refund Attacks and Transaction Malleability}
In~\cite{PatrickMcCorry2016}, modifications are proposed in the \textit{Payment Request} message by adding information about the customer such as registered e-mail address, delivery address, and product information. The payment address should be unique with each Payment Request. Each request is associated with a key, and the same key is used for a refund. However, the use of the additional information poses a threat to the customer privacy. The customer is no longer involved in the information broadcast about the transaction, but the responsibility is to handover the refund to the merchant. Hence all the nodes will learn about the transaction during verification phase and can identify the attacker easily. In particular, the idea is to provide the merchant, a set of publicly verifiable evidence which can cryptographically prove that the refund address received during the protocol belongs to the same pseudonymous customer who authorized the payment.

\par In~\cite{matee2015}, authors propose a manual intervention process that checks the withdrawal transactions to detect a possible malleability attack. Any suspicious pending transactions in the blocks can be seen as a sign of the attack. In addition, all the transactions in the Bitcoin network should have confirmations. In~\cite{AndrychowiczMar2015}, authors show a case of malleability attack on ``deposit protocol'', and provides a solution namely \emph{new deposit protocol}. Finally, the new Segregated Witness~\footnote{$https://en.bitcoin.it/wiki/Segregated_Witness$} (SegWit) proposal stores transaction signatures in a separate merkle tree, prevent unintended transaction malleability, moreover it further enables advanced second-layer protocols such as the Lightning Network, MAST, atomic swaps, and more. Recently. the SegWit soft fork has been activated on the Bitcoin network. More specifically, the SegWit activation means that Bitcoin’s block size limit is replaced by a block ``weight'' limit, which allows for blocks to the size of 4 MB instead of 1 MB.

\subsubsection{Reducing Delays in Processing and Propagation of Transactions} 
In practice, the transactions with a large number of bitcoins are not usually carried out due to the risk of losing it or fear of fraudulent activities. Such transactions are broken into a set of smaller transactions. However, this eventually increases the delay in completing the transaction because the network has to validate more number of transactions. Therefore to reduce this delay, authors in~\cite{Deckerchr2015} suggest performing the payments offline through a separate type of transactions called ``micropayments''~\cite{Rosenbergrg2010} and via a separate channel called micropayment channel. This channel is not a separate network but part of Bitcoin network itself. In a traditional Bitcoin network, users broadcast their transaction and the miners verify it. This happens for all the transactions and the network might get clogged at places where a large number of transaction exists. Also, in such situations, the network gives preference to transactions with large denomination and transaction fees as compared to the smaller ones. Hence, by establishing micropayment channels, the separate dedicated channel is allocated for the counter-parties to perform the transaction. The basic idea is that the transaction is not revealed until both the parties trust each other on their balances and transactions that they wants to perform. If either of the ones misbehaves, then the transaction is broadcasted for the verification in the Bitcoin network. The channels obey the Bitcoin protocols and they are established like any other naive network routing techniques. Hence, these micro payment channels constitute a ``lightning network'' within the Bitcoin network. The advantages of using such a lightning network are as follows:
\begin{itemize}
\item{The technique provides high-speed payments, eliminates the dependency on the third party to validate, reduced load on the Bitcoin network, channels can stay open indefinitely for the transactions, counter-parties can move out of the agreement whenever they want, parties can sign using multiple keys.}
 \item{Parties can broadcast their information when they want for seeking the interference of the other miners to solve the discrepancies.}
 \item{Parties can send their transaction over the channel without revealing their identities to the network and the nodes helping in routing.}
\end{itemize}

\par Transactions propagation delay in Bitcoin network facilitates the double spending attack. Hence accelerating the transaction propagation will help to reduce the probability of performing a successful double spending attack. Authors in~\cite{sallal2017} propose a Bitcoin Clustering Based Ping Time protocol (BCBPT) to reduce the transaction propagation delay by using the proximity information (e.g., ping latencies) while connecting to peers. Moreover, in the context of the selfish mining attack, authors in~\cite{GOBEL201623} study the effect of communication delay on the evolution of the Bitcoin blockchain.

\par In~\cite{GervaisArt2015}, author's provide solutions for \emph{tampering attacks}. A node can announce the time it takes to mine a block together with the advertisement of a new block. This makes another peer in the network to approximately estimate the average time needed to mine a block, and hence no one can spoof by adding unnecessary delays or tampering timestamps. Instead of static timeouts, dynamic timeouts can make more sense since mining time can vary from node to node. All the senders buffer the IP addresses to which it is connecting every time, and this avoids the IP sending same advertise messages again and again to the same peer. A track of all the nodes has to be recorded at every sender and pattern can be analyzed. If a transaction is not replied by a node in a time window, then the sender could ask other nodes to confirm the transaction.

\par Despite all the security threats and their solutions that we have discussed, the number of honest miners in the network is a factor of consideration. More the miners, more people to verify the transactions, hence faster the block validation process and more efficient and secure the consensus process. As the miners are incentive driven, the reward bitcoins can pull more miners into the process, but at the same time the reward reduces half for every four years, hence the miners might migrate towards other cryptocurrencies which offer them higher rewards. 


\par The security issues in Bitcoin are closely linked with the transaction privacy and user anonymity. In Bitcoin right now the users are not really anonymous. The systematic monitoring of the Bitcoin's unencrypted peer-to-peer network and analysis of the public blockchain can reveal a lot of information such as who is using Bitcoin and for what purposes. Additionally, the use of \textit{Know Your Customer} (KYC) policies and \textit{Anti-Money Laundering} (AML) regulation with network traffic and blockchain analysis techniques, could further enhance the quality of the extracted information. From privacy as well as business perspectives, this is not good. For instance, users might not necessarily want the world to know where they spend their bitcoins, how much they own or earn. Similarly, the businesses may not want to leak transaction details to their competitors. Furthermore, the fact that the transaction history of each bitcoin is traceable puts the fungibility of all bitcoins at risk. To this end, we discuss the threats and their existing countermeasures for enabling privacy and enhancing anonymity for Bitcoin users in the following section.

\begin{figure}[h]
\centering
\includegraphics[scale = 0.50]{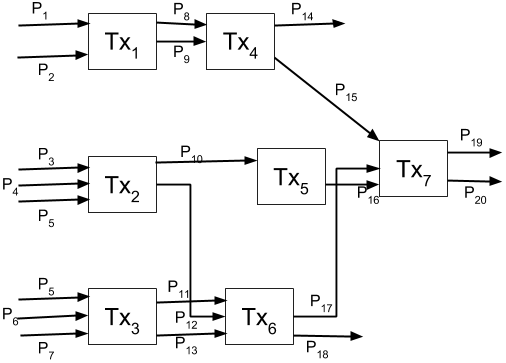}
  \caption{{Blockchain analysis - Transaction graph}}
  \label{Fig:Bitcoin_graph_a}
\end{figure}

\begin{figure}[h]
\centering
\includegraphics[scale = 0.50]{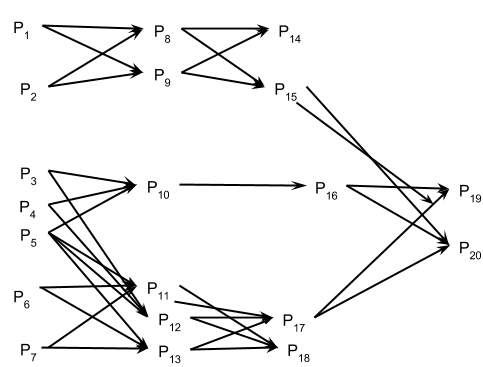}
  \caption{{Blockchain analysis - Address graph}}
  \label{Fig:Bitcoin_graph_b}
\end{figure}

\begin{figure}[h]
\centering
\includegraphics[scale = 0.50]{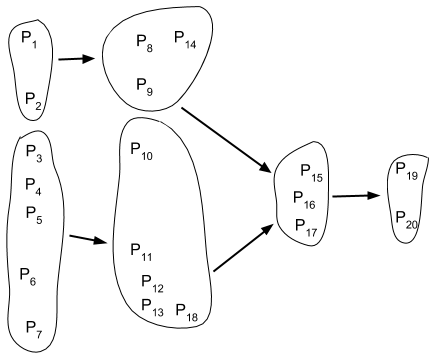}
  \caption{{Blockchain analysis - Entity/User graph}}
  \label{Fig:Bitcoin_graph_c}
\end{figure}


\section{Privacy and Anonymity in Bitcoin}
\label{sec:Bit_PandA}

The traditional banking system achieves a level of privacy by limiting access to transactions information to the entities involved and the trusted third party. While in Bitcoin, the public blockchain reveals all the transaction data to any user connected to the network. However, the privacy can still be maintained upto certain level by breaking the flow of information somewhere in the Bitcoin transaction processing chain. Bitcoin achieves it by keeping public keys anonymous, i.e., the public can see that someone is sending an amount to someone else, but without information linking the transaction to anyone. To further enhance the user privacy, it is advised to use a new key pair for each transaction to keep them from being linked to a particular user. However, linking is still possible in multi-input transactions, which necessarily reveal that their inputs were owned by the same owner. Also, if the owner of a key is revealed, there is a risk that linking could reveal other transactions belonging to the same user. In particular, Bitcoin offers a partial unlinkability (i.e., pseudonymity), and thus it is possible to link a number of transactions to an individual user by tracing the flow of money through a robust blockchain analysis procedure. Bitcoin technology upholds itself when it comes to the privacy, but the only privacy that exists in Bitcoin comes from pseudonymous addresses (public keys or their hashes) which are fragile and easily compromised through different techniques such as Bitcoin address reuse, ``taint'' analysis and tracking payments via blockchain analysis methods, IP address monitoring nodes, web-spidering, to name a few. Once broken, this privacy is difficult and sometimes costly to recover. In~\cite{Koshy2014} authors highlight the fact that the Bitcoin does not have any directory to maintain the log and other transaction-related information. However, an adversary can associate the offline data such as emails and shipping addresses with the online information, and it can get the private information about the peers. In this section, we discuss the various security threats to privacy and anonymity of the Bitcoin users and the corresponding state-of-the-art solutions that are proposed to enhance the same.

\subsection{Blockchain Analysis and Deanonymization}
\label{sec:Bit_PandA_Txgraph}

A complete anonymity in Bitcoin is a complicated issue. To enforce anonymity in transactions, the Bitcoin allows users to generate multiple Bitcoin addresses and it only stores the mapping information of a user to her Bitcoin addresses on the user's device. As a user can have multiple addresses, hence an adversary who is trying to deanonymize needs to construct a one-to-many mapping between the user and its associated addresses. In particular, the Bitcoin users can be linked to a set of public addresses by using a detailed blockchain analysis procedure~\cite{Rondo2013}. Authors in~\cite{Koshy2014} show that the two non-trivial networking topologies called \textit{transaction network} and \textit{user network}, which provides reciprocal views of the Bitcoin network and have possible adverse implications for user anonymity. Similar to the work done in~\cite{Koshy2014}, authors in~\cite{Androulakielli2013} presents an evaluation for privacy concerns in Bitcoin by analyzing the public blockchain. The analysis of blockchain requires three pre-processing steps, which includes: 

\begin{itemize}
\item \textit{Transaction graph}: The whole blockchain could be viewed as an acyclic \textit{transaction graph} $G_t = \{T, E\}$, where $T$ is a set of transactions stored in the blockchain, and $E$ is the set of unidirectional edges between these transactions. A $G_t$ represents the flow of bitcoins between transactions in the blockchain over time. The set of input and output bitcoins in a transaction can be viewed as the weights on the edges in a $G_t$. In particular, each incoming edge $e \in E$ in a transaction carries a timestamp and the number of bitcoins ($C_i$) that forms an input for that transaction. Figure~\ref{Fig:Bitcoin_graph_a} shows an instance of transaction graph for a set of transactions stored in the blockchain.

\item \textit{Address graph}: By traversing the transaction graph we can easily infer the relationship between various input and output Bitcoin addresses, and using these relations we can generate an \textit{address graph}, $G_a = \{P, E'\}$, where $P$ is the set of Bitcoin addresses and $E'$ are the edges connecting these addresses. Figure~\ref{Fig:Bitcoin_graph_b} shows an address graph derived from Figure~\ref{Fig:Bitcoin_graph_b}.  

\item \textit{User/entity graph}: By using the address graph along with a number of heuristics which are derived from Bitcoin protocols, the next step is to create an \textit{entity graph} by grouping addresses that seem to be belonging to the same user. The entity graph, $G_e = \{U, E''\}$, where $U$ is a disjoint subset of public keys ($p$) such that $p \in P$ and $E''$ are the edges connecting different $U's$ to show a directed connectivity between them. Figure~\ref{Fig:Bitcoin_graph_c} shows the entity graph derived from Figure~\ref{Fig:Bitcoin_graph_b} based on a set of heuristics. 

\end{itemize} 

\par In~\cite{Androulakielli2013}, authors introduce two heuristics that are derived directly from Bitcoin protocols or its common practices. The first is the most widely used heuristic that provides an adequate level of linkability and it heavily depends on the implementation details of Bitcoin protocols, and are termed as \textit{idioms of use} as mentioned in~\cite{Meiklejohny2013}. The \textit{idioms of use} assumes that all the inputs in a transaction are generated by the same user because in practice different users rarely contribute in a single, collaborative transaction. This heuristic also supports the fact that transitive closure can be applied to the transaction graph to yield clusters of Bitcoin addresses. For instance, by applying the above heuristic along with its transitive property on Figure~\ref{Fig:Bitcoin_graph_a}, one can assume that transactions $Tx_2$ and $Tx_3$ are initiated by the same user as both shares a common input $p_5$, hence the addresses ranging from $p_3$ to $p_6$ could belong to the same user. The second heuristic links the input addresses of a transaction to its output addresses by assuming that these outputs as \textit{change} addresses if an output address is completely new (i.e., the address has never appeared in the past and it will not be seen in the blockchain to be re-used to receive payments). In Figure~\ref{Fig:Bitcoin_graph_b}, the addresses $p_{14}$ and $p_{18}$ satisfy the second heuristic, and thus these addresses can be clustered with their inputs as shown in the Figure~\ref{Fig:Bitcoin_graph_c}. Authors in~\cite{Meiklejohny2013} argued that the aforementioned heuristics are prone to errors, in cases where the implementation of Bitcoin protocols change with time, and the traditional Bitcoin network also changes which now consists of more number of mining pools instead of single users. Due to these facts, it is possible that the entity graph might contain a large number of false positives in the clustering process, hence it leads to the further refinements in the above heuristics. To reduce the false positives, authors in~\cite{Meiklejohny2013} suggest the manual inspection process to identify the usage patterns induced by Bitcoin services (such as SatoshiDice). For instance, SatoshiDice requires that the payouts use the same address, therefore if a user spent coins using a change address, the address would receive another input which invalidates the one-time receive property of a change address. Furthermore, in~\cite{StevenGoldfederr2016} authors exploit the multi-signature addressing technique for the purpose of adverse effect on the user privacy. Authors conclude that even if the Bitcoin addresses are changed, the structure of the \textit{change} address in a multi-signature transaction can be matched to its input addresses.       

\par Apart from using the adaptable and refined heuristics to match with the constantly changing blockchain usage patterns and Bitcoin services, the adversary needs to take further steps to link the address clusters with the real-world identities once an entity graph with low false positives is created. Authors in~\cite{Meiklejohny2013} perform with high precision the linking of clusters with the online wallets, vendors, and other service providers as one can do several interactions with these entities and learn at least one associated address. However, identifying regular users is difficult with the same approach, but the authors also suggest that authorities with subpoena power might even be able to identify individual users since most of the transaction flow passes through their centralized servers. These servers usually require keeping records for customer identities. Furthermore, the use of side-channel information is considered helpful in mapping the addresses. For instance, WikiLeaks, Silk Road, to name a few, uses publicly known addresses, and many service providers such as online sellers or exchange services require the user identity before providing a service. One can also make use of the web crawlers (such as bitcointalk.org) that searches the social networks for Bitcoin addresses~\cite{micFlederKP15}~\cite{Reidfer2013}. 

\par A commercial approach for blockchain analysis could be to use the software BitIodine~\cite{Spagnuoloo2014} that offers an automated blockchain analysis framework. Due to its rapid growth in such a short span of time, the Bitcoin networks has become of great interest to governments and law enforcement agencies all over the world to track down the illicit transactions. By predicting that there is a huge market potential for Bitcoin, various companies such as Elliptic, Chainalysis, Numisight, Skry, to name a few, are specializing in ``bitcoin blockchain analysis'' models. These companies provide a set of tools to analyze the blockchain to identify illicit activities and even help to identify the Bitcoin users in the process. Authors in~\cite{DBattista2015} propose \textit{BitConeView}, a graphical tool for the visual analysis of bitcoins flow in a blockchain. BitConeView allows to graphically track how bitcoins from the given sources (i.e., transaction inputs) are spent over time by means of transactions and are eventually stored at multiple destinations (i.e., unspent transaction outputs). 

\par Recently, authors in~\cite{Goldfederde2017} analyze the impact of online tracking on the privacy of Bitcoin users. The paper shows that if a user purchases by paying with cryptocurrency such as bitcoins, an adversary can uniquely identify the transaction on the blockchain by making use of the third-party trackers which typically possess enough information about the purchase. Latter, these transactions could be linked to the user cookies and then with the real identity of a user, and user's purchase history is revealed. Furthermore, if the tracker is able to link the two purchases of the same user to the blockchain in this manner, it can identify the user's entire cluster of Bitcoin addresses and transactions on the blockchain through the use of standard tracking software and blockchain analysis techniques. The authors show that these attacks are resilient against the existing blockchain anonymity techniques such as \textit{CoinJoin}~\cite{GmMaxwell2013}. Also, these attacks are passive, hence can be retroactively applied to past purchases as well.

\par Finally, network de-anonymization could be used to link an IP address to a user in the Bitcoin's P2P network because while broadcasting a transaction the node leaks its IP address. Same as the blockchain analysis, a rigorous way to link IP addresses to hosts is by exploiting the network related information that can be collected by just observing the network. Over the years, multiple deanonymization attacks in which an adversary uses a ``supernode'' that connects with the active peers and listen to the transaction traffic relayed by honest nodes in the network~\cite{Biryukov2014}~\cite{AbBiryukov2015}~\cite{Koshy2014} are proposed. By exploiting the symmetric diffusion of transactions over the network, it is possible to link the Bitcoin users' public keys to their IP addresses with an accuracy of nearly 30\%~\cite{Biryukov2014}. Moreover, the use of ``supernode'' for linking is trivial, hence it exploits only minimal knowledge of the P2P graph structure and the structured randomness of diffusion. Therefore, we can hypothesize that even higher accuracies could be achieved by using more sophisticated network traffic analyzing techniques.

\begin{figure*}[h]
\centering
\includegraphics[scale = 0.40]{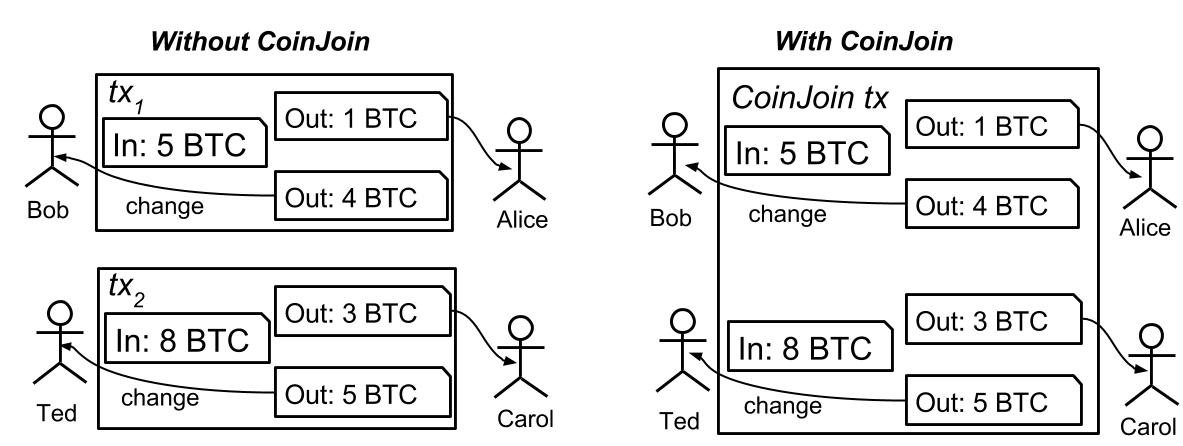}
  \caption{Example: CoinJoin basic idea}
  \label{fig:CoinJoin}
\end{figure*}

\begin{table*}
\large
\caption {Techniques for Improving Privacy and Anonymity in Bitcoin}
\centering
\scalebox{0.70}{
\begin{tabular}{|p{3.5cm}|p{2cm}|p{6cm}|p{6cm}|p{6cm}|}\hline
\textbf{Proposals} & \textbf{Type/Class} & \textbf{Distinct features and properties} & \textbf{Advantages} & \textbf{Disadvantages}\\ \hline

\textit{CoinJoin}~\cite{GmMaxwell2013} & P2P & uses multi-signature transactions to enhance privacy & prevent thefts, lower per-transaction fee & anonymity level depends on the number of participants, vulnerable to DoS (by stalling joint transactions), sybil and intersection attacks, prevents plausible deniability \\\hline
\textit{CoinShuffle}~\cite{RuffingTim2014} & P2P & decentralized protocol for coordinating CoinJoin transactions through a cryptographic mixing protocol & internal unlinkability, robust to DoS attacks, theft resistance & lower anonymity level and deniability, prone to intersection and sybil attacks \\\hline
\textit{Xim}~\cite{Bissias2014} & P2P & anonymously partnering and multi-round mixing & distributed pairing, internal unlinkability, thwarts sybil and DoS attacks & higher mixing time \\\hline
\textit{CoinShuffle++ / DiceMix}~\cite{TimRuff2016} & P2P & based on CoinJoin concept, optimal P2P mixing solution to improve anonymity in crypto-currencies & low mixing time (8 secs for 50 peers), resistant to deanonymization attack, ensures sender anonymity and termination & vulnerable to DoS and sybil attacks, limited scalability, no support for Confidential Transactions (CT) \\\hline
\textit{ValueShuffle}~\cite{TimRuff2017} & P2P & based on CoinShuffle++ concept, uses Confidential Transactions mixing approach to achieve comprehensive transaction privacy & unlinkability, CT compatibility and theft resistance, normal payment using ValueShuffle needs only one transaction & vulnerable to DoS and sybil attacks, limited scalability \\\hline
\textit{Dandelion}~\cite{Venkatakrishnan17} & P2P & networking policy to prevent network-facilitated deanonymization of Bitcoin users & provides strong anonymity even in the presence of multiple adversaries & vulnerable to DoS and sybil attacks \\\hline
\textit{SecureCoin}~\cite{Ibrahimg2017} & P2P & based on CoinParty concept, an efficient and secure protocol for anonymous and unlinkable Bitcoin transactions & protect against sabotage attacks, attempted by any number of participating saboteurs, low mixing fee, deniability & vulnerable to DoS attacks, limited scalability\\\hline
\textit{CoinParty}~\cite{Ziegeldor2015} & partially P2P & based on CoinJoin concept, uses threshold ECDSA and decryption mixnets to combine pros of centralized and decentralized mixes in a single system & improves on robustness, anonymity, scalability and deniability, no mixing fee & partially prone to coin theft and DoS attack, high mixing time, requires separate honest mixing peers \\\hline

\textit{MixCoin}~\cite{BonneauJos2014} & Distributed & third-party mixing with accountability & DoS and sybil resistance & partial internal unlinkability and theft resistance, \\\hline
\textit{BlindCoin}~\cite{Valentaluke2015} & Distributed & based on MixCoin concept, uses blind signature scheme to ensure anonymity & internal unlinkability, DoS and sybil resistance & partial theft resistance, additional costs and delays in mixing process\\\hline
\textit{TumbleBit}~\cite{EthanHeilman2016} & Distributed & undirectional unlinkable payment hub that uses an untrusted intermediary & prevents theft, anonymous, resists intersection, sybil and DoS, scalable (implemented with 800 users) & normal payment using TumbleBit needs at least two sequential transactions\\\hline

\end{tabular}}
\label{T:mixing}
\end{table*}

\subsection{Proposals for enabling privacy and improving anonymity}
\label{sec:Bit_PandA_enabling}
 
Privacy is not defined as an inherent property in Bitcoin initial design, but it is strongly associated with the system. Therefore, in recent years an array of academic research~\cite{Androulakielli2013}~\cite{Barbersimonn2012}~\cite{Meiklejohns2015}~\cite{Spagnuoloo2014} which shows various privacy-related weaknesses in the current Bitcoin protocol(s) has been surfaced. This research triggered a large set of privacy-enhancing technologies~\cite{Barbersimonn2012}~\cite{BSasson2014}~\cite{Bissias2014}~\cite{BonneauJos2014}~\cite{EthanHeilman2016}~\cite{Heilmaneth2016}~\cite{TimRuff2016}~\cite{Ziegeldor2015} aiming at strengthening privacy and improving anonymity without breaking Bitcoin fundamental design principles. In this section, we discuss these state-of-the-art protocols which work toward the enhancement of privacy and anonymity in Bitcoin.  

\par Based on the aforementioned discussion in Section~\ref{sec:Bit_PandA}, it is evident that the public nature of the blockchain poses a significant threat to the privacy of Bitcoin users. Even worse, since funds can be tracked and tainted, no two coins are equal, and fungibility, a fundamental property required in every currency, is at risk. With these threats in mind, several privacy-enhancing technologies have been proposed to improve transaction privacy in Bitcoin. The state-of-the-art proposals (refer tables~\ref{T:mixing} and~\ref{T:altcoins}) for enabling privacy in cryptocurrencies can be broadly classified into three major categories namely, \textit{Peer-to-peer mixing protocols}, \textit{Distributed mixing networks}, and \textit{Altcoins}. 

\subsubsection{Peer-to-peer mixing protocols}
\label{sec:mixing}

Mixers are anonymous service providers, that uses mixing protocols to confuse the trails of transactions. In mixing process, the client's funds are divided into smaller parts. These parts are mixed at random with similar random parts of other clients, and you end up with completely new coins. This helps to break any link between the user and the coins she purchased. However, mixers are not an integral part of Bitcoin, but various mixing services are heavily used to enhance the anonymity and unlinkability in the system. In peer-to-peer (P2P) mixing protocols~\cite{CorriganGibbs2010}~\cite{Ibrahimg2017}~\cite{RuffingTim2014}, a set of untrusted Bitcoin users simultaneously broadcast their messages to create a series of transactions without requiring any trusted third party. The main feature of a P2P mixing protocol is to ensure sender anonymity within the set of participants by permuting ownership of their coins. The goal is to prevent an attacker which controls a part of the network or some of the participating users to associate a transaction to its corresponding honest sender. The degree of anonymity in P2P protocols depends on the number of users in the anonymity set.

\par Table~\ref{T:mixing} shows a range of P2P mixing protocols along with their brief description, advantages, and disadvantages in terms of user anonymity and transaction security. CoinJoin~\cite{GmMaxwell2013}, a straightforward protocol for implementing P2P mixing, it aims to enhance privacy and securely prevent thefts. Figure~\ref{fig:CoinJoin} shows CoinJoin basic idea with an example in which two transactions (i.e., $tx_1$ and $tx_2$) are joined into one while inputs and outputs are unchanged. In CoinJoin, a set of users with agreed (via their primary signatures) inputs and outputs create a standard Bitcoin transaction such that no external adversary knows which output links with which input, hence it ensures external unlinkability. To prevent theft, a user only signs a transaction if its desired output appears in the output addresses of the transaction. In this way, CoinJoin makes the multiple inputs of a transaction independent from each other, thus it breaks the basic heuristic from Section~\ref{sec:Bit_PandA_Txgraph} (i.e., inputs of a transaction belong to the same user). However, CoinJoin has few major drawbacks, which includes limited scalability and privacy leakage due to the need of managing signatures of the involved participants in the mixing set, the requirement of signing a transaction by all its participants make CoinJoin vulnerable to DoS attacks, and to create a mix each participant has to share their signature and output addresses within the participating set which causes internal unlinkability. To address the internal unlinkability issue and to increase the robustness to DoS attacks, authors in~\cite{RuffingTim2014} propose CoinShuffle, a decentralized protocol that coordinates CoinJoin transactions using a cryptographic mixing technique. Later, an array of protocols~\cite{TimRuff2016}~\cite{TimRuff2017}~\cite{Ibrahimg2017} are built on the concept of either CoinJoin or CoinShuffle that enhances the P2P mixing by providing various improvements, that includes resistance to DoS, sybil, and intersection attacks, plausible deniability, low mixing time, and scalability of the mixing groups.

\subsubsection{Distributed mixing networks}
\label{sec:Dist_mixing}

Authors in~\cite{BonneauJos2014} propose \textit{MixCoin}, a third-party mixing protocol to facilitate anonymous payments in Bitcoin and similar cryptocurrencies. The \textit{MixCoin} uses the emergent phenomenon of currency mixes, in which a user shares a number of coins with a third-party mix using a standard-sized transaction, and it receives back the same number of coins from the mix that is submitted by some other user, hence it provides strong anonymity from external entries. \textit{MixCoin} uses a reputation-based cryptographic accountability technique to prevent other users within the mix from theft and disrupting the protocol. However, mixes might steal the user coins at any time or become a threat to the user anonymity because the mix will know the internal mapping between the users and outputs. To provide internal unlinkability (i.e., preventing the mix from learning input-output linking) in \textit{MixCoin}, authors in~\cite{Valentaluke2015} proposes \textit{BlindCoin} which extends the \textit{MixCoin} protocol by using blind signatures to create user inputs and cryptographically blinded outputs called \textit{blinded tokens}. However, to achieve this internal unlinkability, \textit{BlindCoin} requires two extra transactions to publish and redeem the blinded tokens, and the threat of theft from the mix is still present. 

\par Recently, in~\cite{EthanHeilman2016} authors propose \textit{TumbleBit}, a Bitcoin-compatible unidirectional unlinkable payment hub that allows peers to make fast, off-blockchain payments anonymously through an untrusted intermediary called \textit{Tumbler}. Similar to Chaumian original eCash protocol~\cite{Chaum1983}, TumbleBit enforces anonymity in the mixing by ensuring that no one, not even the Tumbler, can link a transaction of its sender to its receiver. The mixing of payments from 800 users shows that TumbleBit provides strong anonymity and theft resistance and it is scalable.                      

\begin{table*}
\large
\caption {Summary of Altcoins}
\centering
\scalebox{0.70}{
\begin{tabular}{|p{3.5cm}|p{6cm}|p{6cm}|p{6cm}|}\hline
\textbf{Proposals} & \textbf{Distinct features and properties} & \textbf{Advantages} & \textbf{Disadvantages}\\ \hline

\textit{ZeroCoin / ZeroCash / Zcash}~\cite{Mierscc2013}~\cite{BSasson2014} & a cryptographic extension to Bitcoin , unlinkable and untraceable transactions by using zero knowledge proofs & provides internal unlinkability, theft and DoS resistance & relies on a trusted setup and non-falsifiable cryptographic assumptions, blockchain pruning is not possible\\\hline
\textit{CryptoNote}~\cite{SaberhagenN2013} & relies on ring signatures to provide anonymity & provides strong privacy and anonymity guarantees & higher computational complexity, not compatible with pruning \\\hline
\textit{MimbleWimble}~\cite{JedusorT2016} ~\cite{Poelstra2016} & a design for a cryptocurrency with confidential transactions & CT compatibility, improve privacy, fungibility and scalability & vulnerable to DoS attacks, not compatible with smart contracts\\\hline
\textit{ByzCoin}~\cite{Eleftherios2016} & Bitcoin-like cryptocurrency with strong consistency via collective signing & lower consensus latency and high transaction throughput, resistance to selfish and stubborn mining~\cite{Nayakkk2016}, eclipse and delivery-tampering and double-spending attacks & vulnerable to slow down or temporary DoS attack and 51\% attack,  \\\hline
\textit{Ethereum (ETH)~\cite{ether2014}} & uses proof-of-stake, open-ended decentralized software platform that enables Smart Contracts and Distributed Applications & run without any downtime, fraud, control or interference from a third party, support developers to build and publish distributed applications & scalability issues (uses complex network), running untrusted code, limited (i.e., non-turing-complete) scripting language  \\\hline
\textit{Mastercoin (or Omni)~\cite{Willett2013}} & uses enhanced Bitcoin Core ad Proof of Authenticity, Colored coins, Exodus address & Easy to use, secure web wallets available, Escrow fund (insurance against panic), Duress protection using a trusted entity & wallets handling the transactions should aware of the concept of colored coins, possibility to accidentally uncolor colored coin assets exists\\\hline
\textit{Litecoin (LTC, litecoin.org)} & uses Segwit, which allows technologies like Lightning Network & scalable, low transaction mining time, anonymous and cheaper & very few stores accept payment in Litecoins, high power consumption \\\hline
\textit{Dash} (DASH, dashpay.io) & uses Proof of Service, implements native CoinJoin like transactions & higher privacy (mixes transactions using master nodes), InstantX provides faster transaction processing  & less liquid,technology is too young, does not yet have a critical mass of merchants or users \\\hline
\textit{Ripple (XRP, ripple.com)} & implements a novel low-latency consensus algorithm based on byzantine agreement protocol & fast transaction validation, less energy-intensive, no 51\% attack & not fully decentralized, vulnerable to attacks such as consensus split, transaction flood and software backdoor  \\\hline
\textit{Monero (XMR, getmonero.org)} & based on the CryptoNote protocol,  & improves user privacy by using  ring signatures, lower transaction processing time (average every 2 minutes)  & transaction linkability could be achieved by leveraging the ring signature size of zero, output merging, temporal analysis  \\\hline
\textit{Counterparty (XCP, counterparty.io)} & created and distributed by destroying bitcoins in a process known as \textit{proof of burn} & same as bitcoins & same as bitcoins \\\hline

\end{tabular}}
\label{T:altcoins}
\end{table*}

\subsubsection{Bitcoin extensions or Altcoins}
\label{sec:Dist_mixing}

\par Bitcoin has not just been a most popular cryptocurrency in today's market, but it ushers a wave of other cryptocurrencies that are built on decentralized peer-to-peer networks. In fact, the Bitcoin has become the de facto standard for the other cryptocurrencies. The other currencies which are inspired by Bitcoin are collectively known as \textit{altcoins}. Instead of proposing techniques (such as mixing and shuffling) to increase transaction anonymity and user privacy, the altcoins work as an extension to Bitcoin or a full-fledged currency. The popular altcoins along with their brief description have been shown in Table~\ref{T:altcoins}. Some of these currencies are easier to mine than Bitcoin however, there are tradeoffs, including greater risk brought on by lesser liquidity, acceptance, and value retention. 

\par Authors in~\cite{Mierscc2013} propose \textit{ZeroCoin}, a cryptographic extension to Bitcoin which provides anonymity by design by applying zero-knowledge proofs which allow fully encrypted transactions to be confirmed as valid. It is believed that this new property could enable entirely new classes of blockchain applications to be built. In ZeroCoin, a user can simply wash the linkability traces from its coins by exchanging them for an equal value of ZeroCoins. But unlike the aforementioned mixing approaches, the user should not have to ask for the exchange to a mixing set, instead, the user can itself generate the ZeroCoins by proving that she owns the equal value of bitcoins via the Zerocoin protocol. For instance, $Alice$ can prove to others that she owns a bitcoin and is thus eligible to spend any other bitcoin. For this purpose, first, she produces a secure commitment, i.e., the zerocoin, which is recorded in the blockchain so that others can validate it. In order to spend a bitcoin, she broadcasts a zero-knowledge proof for the respective zerocoin, together with a transaction. The zero-knowledge cryptography protects $Alice$ from linking the zerocoin to her. Still, the other participants can verify the transaction and the proof. Instead of a linked list of Bitcoin transactions, Zerocoin introduces intermediate steps. In this way, the use of zero-knowledge proofs prevent the transaction graph analyses. Unfortunately, even though Zerocoin’s properties may seem appealing, it is computationally complex, bloats the blockchain and requires protocol modifications. However, it demonstrates an alternative, privacy-aware approach. Currently, ZeroCoin derives both its anonymity and security against counterfeiting from strong cryptographic assumptions at the cost of substantially increased computational complexity and size. 

\par An extension of ZeroCoin called \textit{ZeroCash} (also know as Zcash) is presented by~\cite{BSasson2014}. ZeroCash uses an improved version of zero-knowledge proof (in terms of functionality and efficiency) called \textit{zk-SNARKs}, which hides additional information about transactions such as the amount and recipient addresses to achieve strong privacy guarantees. However, ZeroCash relies on a trusted setup for generation of secret parameters required for SNARKs implementation, it requires protocol modifications, and the blockchain pruning is not possible. Recently, authors in~\cite{JedusorT2016} propose \textit{MimbleWimble}, an altcoin that supports confidential transactions (CT). The CTs can be aggregated non-interactively and even across blocks, thus greatly increases the scalability of the underlying blockchain. However, such aggregation alone does not ensure input-output unlinkability against parties who perform the aggregation, e.g., the miners. Additionally, Mimblewimble is not compatible with smart contracts due to the lack of script support. 

\par Beyond Bitcoin, the so-called second generation of cryptocurrencies, such as Ethereum (Ether), Mastercoin (MSC), Counterparty (XCP) are introduced in the market. These cryptocurrencies implement a new transaction syntax with a fully-fledged scripting language written in Turing complete language. Furthermore, these cryptocurrencies understood the digital assets in terms of smart contracts and colored coins. Unlike Bitcoin, Ethereum was designed to be much more than a payment system. In particular, it is a decentralized platform that runs smart contracts, which are the applications that run exactly as programmed without any possibility of downtime, censorship, fraud or third-party interference. This implies that these digital assets can be used to realize sophisticated financial instruments such as stocks with automatic dividend payouts or to manage and trade physical properties such as a house. Most of these next-generation coins work on top of Bitcoin’s blockchain and are therefore also known as \textit{on-chain currencies}. Since they encode their transactions into Bitcoin’s transactions, they lack the validation of transactions by miners, because Bitcoin miners do not \textit{understand} the new transaction types. For this purpose, a new protocol layer is built upon Bitcoin’s strong foundation and its security. Furthermore, it is seen as an increase in Bitcoin’s value from which both will profit. As a detailed discussion on the altcoins is out of the scope of our work, we direct interested readers to the existing literature such as~\cite{FTschorsch2015} and~\cite{Mukho2016}.


\par As a summary, in this section, the Bitcoin's privacy and anonymity concerns have been discussed. It is observed that Bitcoin is pseudo-anonymous as the account is tied to the random and multiple Bitcoin addresses and not to the individual users. With the rapidly increasing popularity of bitcoins, the need for privacy and anonymity protection also increases, and it must be ensured that the users will receive a satisfactory level of service in terms of privacy, security, and anonymity.

\section{Summary of observations and future research directions}
\label{sec:future_work}

After our comprehensive survey on the security and privacy aspects of Bitcoin and its major related techniques, we now summarize our lessons learned, before presenting the possible future challenges and research directions. Some of these are already discussed in previous sections. However, remaining challenges and open research issues are dealt in brief in this section.

\par With the use of proof-of-work based consensus algorithm and a secure timestamping service, Bitcoin provides a practical solution to the Byzantine Generals problem. However, to achieve distributed consensus, Bitcoin exposes itself to a number of security threats. The main threat is double spending (or race attacks) which will always be possible in the Bitcoin. The transparency in the system is provided by using an unforgeable distributed ledger (i.e., blockchain), which holds all the transactions ever processed, in such a way that anyone can verify their integrity and authenticity. But, at the same time, this transparency introduces a ubiquitous global attacker model. Hence, we can be deduced from the discussion presented in Section~\ref{sec:Bit_PandA} that Bitcoin is anything but private. Nevertheless, Bitcoin provides pseudonymity by hiding identities and the research community is putting a lot of efforts to further strengthen this property. For instance, use of commitment schemes such as zero-knowledge proofs greatly improves unlinkability and untraceability in transactions.

\par One of the major contribution of Bitcoin is the degree of transparency and decentralization, that it provides along with the adequate level of security and privacy, which was previously deemed impossible. The original concept of mining, which could be based on proof of work, proof of stake, proof of burns or some other scheme, not only secures the blockchain but it eventually achieves the distributed consensus. In particular, the most important steps that make the whole process so cohesive includes, the way these schemes binds the \textit{votes} to something \textit{valuable}, give rewards in exchange to \textit{pay} for these valuables, and at the same time controls the supply of the cryptocurrencies in the system. Without these mining schemes, the fake identities would be able to easily disturb (through sybil attack) the consensus process and destroy the system. Due to this, i.e., availability of a mining based consensus protocol, we can safely conclude that 51\% attacks are the worst case scenario for Bitcoin. However, the rapidly increasing mining pools threatens the decentralization of Bitcoin.

\par In the near future it is hard to comment on the survivability of the Bitcoin, i.e., whether Bitcoin can and will stay as robust as it is today. In particular, the scalability of the network, the continuously decreasing rewards, increasing transaction fee, and the security and privacy threats are the pressing issues, which needs to be addressed. The peer-to-peer network already seems to be having the symptoms of degradation, which can be seen in terms of propagation delay for both, the new transaction generated by a user and the newly validated block by a miner. This network propagation delay becomes, particularly a major issue, because the Bitcoin’s security assumptions heavily rely on the fast propagation of transactions and blocks. Therefore, it is very important that the Bitcoin network is easy to scale to more participants and it is able to handle higher transaction rates. In case of the subsiding mining rewards, the research community is unsure whether this poses a real problem or if fees are able to provide the necessary incentive. So far, various improvements and altcoins have been implemented to resolve the aforementioned issues (please refer to tables~\ref{T:mixing} and~\ref{T:altcoins}. However, it remains unclear which of the alternative approaches are most promising in terms of practical implementation that will actually improve Bitcoin.

\par From the improvement perspective, Bitcoin can consider all the altcoins as a huge testing environment, from which it can borrow novel techniques and functionalities to address its weaknesses. At least in the recent future, the Bitcoin will be constantly evolving and will be in the under development phase, hence we now present few research directions that could be exploited in this direction.
\begin{itemize}

\item \textit{Game theory and stability:} Recall that mining pools consist of individual miners who pool their hashing power as well as their incentives. Miners can behave selfishly by holding on to their blocks and releasing it whenever they want. This kind of selfish behavior may pose a game theoretic problem between the selfish miners and the network. Since all the miners perform with a notion of increasing their incentives, a game theoretic approach is well suited for achieving Nash equilibrium among miners (i.e., players)~\cite{AggelosKi16}. Attackers may try to contribute to an increase of their chain length compared to honest chain in the network. This poses a game between the honest chain miners and the malicious miners, thus achieving equilibrium to bring stability in the network is a possible research direction. There are numerous proposals~\cite{AggelosKi16}~\cite{Lewenbergyo2015}~\cite{benFischPS17} which shows that the use of the game-theoretic approaches provide useful information about the effects of selfish mining, block withholding and discarding attacks, and the incentive distribution problem in the mining pools. Therefore, we believe that this approach could be effectively used for modeling the various issues and providing adequate solutions for the identified issues related to the mining pools.     

\item \textit{Cryptographic and keying techniques:} The Simplified Payment Verification (SPV) protocol which is a lightweight protocol used for the verification of the transaction sent from a user~\cite{Kiayiasagg2016}, and it is often vulnerable to attacks like sybil and double spending. A more robust verification protocol is a current requirement. For the key manipulations and calculations, a distributed approach is always preferred more than the centralized one. This is to avoid the point of failure or the central server under the risk of an attack. Hence, in this direction, the innovative means of key computation and storage of the bitcoins in a distributed fashion is a possible research direction. Additionally, the Bitcoin protocols use EDCSA and hash functions like SHA-256 which creates another research scope as there is always an adequate requirement to improve these algorithms or implement novel keying and hashing techniques. We have seen the use of cluster or group keys which are based on some threshold in order to solve various attacks. For instance, fix a group head and get an additional signature or authentication on every transaction~\cite{Barbersimonn2012}. Another approach is to use ``trusted paths'' which is based on hardware that allows users to read and write a few cryptographic data~\cite{Barbersimonn2012}. Finally, there are few techniques which use Bloom filters for securing wallets. Nevertheless, filters might lead to false positives and false negatives that will consume the network bandwidth, thus reducing it can be a potential research directive.

\item \textit{Improving blockchain protocol:} Blockchain provides for the first time a probabilistic solution to the Byzantine Generals problem~\cite{Lamport1982}, where consensus is reached over time (after confirmations) and makes use of economic incentives to secure the functionality of the overall infrastructure. The blockchain technology promises to revolutionize the way we conduct business. For instance, blockchain startups have received more than one billion dollars~\cite{Coindesk14} of venture capital money to exploit this technology for applications such as voting, record keeping, contracts, to name a few. Despite its potential, blockchain protocol faces significant concerns in terms of its privacy~\cite{Herremartí2016} and scalability~\cite{Sompolinsky2015}~\cite{llLewenberg2015}. The append-only nature of the blockchain is essential to the security of the Bitcoin ecosystem as transactions are stored in the ledger forever and are immutable. However, an immutable ledger is not appropriate for all new applications that are being envisaged for the blockchain. Recently, authors in~\cite{GiuseppeAt2016} present modification in blockchain techniques that allows operation such as re-writing one or more blocks, compressing any number of blocks into a smaller number of blocks, and inserting one or more blocks.

\item \textit{Fastness:} Bitcoin's proof of work is designed to validate a new block on average every 10 minutes, and it is recommended to wait for six confirmations before accepting a transaction~\cite{joRosenfeld14}, which makes it impractical for many real-world applications (e.g., a point of sale payments). Faster mining with the same robustness such as one proposed in~\cite{Eleftherios2016} is a future requirement. Recently authors in~\cite{DMilutinovicHWK17} present \textit{Proof of Luck}, an efficient blockchain consensus protocol to achieve low-latency transaction validation, deterministic confirmation time, negligible energy consumption, and equitably distributed mining. 

\item \textit{Incentives for miners:} In general, incentives can be either fixed or variable depending on the complexity of the puzzle that miners solve. A variable incentive may increase the competition between the miners and help to solve puzzles that are challenging. The miners who inform the malfunctions and other illegal behavior in the network can be awarded additional coins as a reward. This act will increase the number of honest nodes in the network. In the world of growing demand for the cryptocurrencies, there is a lot of competition for bitcoins or any other digital currency to retain its popularity in the market. Additionally, miners may migrate by looking at the rewards given by the other competitors or by the fact that for every four years the incentives are halved. Therefore, essential questions that need addressing includes, how to make the miners fix to a currency in such a competitive environment, and what are the other incentives the Bitcoin system can think of to attract the miners.

\item \textit{Smart contracts and preventing backtracks:} Smart contract refers to the computer programs that embody a self-executing and self-enforcing contract to which users may become a party, by interacting with it electronically. These contracts are of particular interest to those in the financial sector. However, the concept of smart contract is not a new one, but the advent of blockchain technology spurred interest in it because the blockchain eliminates the need to rely on a trusted third party to ``execute'' the contract, and enables to use of cryptocurrency as ``programmable money''. Bitcoin’s support for smart contracts is extremely limited. Recently authors in~\cite{AKosba2016} propose \textit{Hawk}, which uses a blockchain model of cryptography to generate privacy-preserving smart contracts. Similar to Bitcoin, authors in~\cite{DBZyskindNP15} proposes \textit{Enigma}, a decentralized computation platform which provides a highly optimized version of secure multi-party computation with guaranteed privacy to effectively execute smart contracts. 

\end{itemize}

\section{conclusions}
\label{sec:conclusion}

Bitcoin has already evinced a popular digital currency in the market. However, the fame of Bitcoin has attracted antagonists to use Bitcoin network for their selfish motives and benefits. Today we have approximately 1146 different cryptocurrencies in action, out of which many are a recent introduction to the market. From all these fiat-currencies, the outstanding popularity and high market capital of bitcoins make it attractive for adversaries to launch various security threats. According to our survey, even though the construction of the Bitcoin protocols with proof-of-work and consensus to protect the user actions are the robust features in Bitcoin, these itself becoming a point of manipulation for cyber thieves. Starting from packet sniffing to the double spending, the Bitcoin is dreaded with various attacks. Though literature provides solutions against few of these attacks, the robust and effective security solutions that can ensure proper functioning of Bitcoin in the future are still absent. Together with security, the distributed nature of Bitcoin blockchain has lead glitches in the privacy and anonymity requirements of the users. In summary, this paper is a sole attempt towards highlighting the security and privacy issues in different fields of Bitcoin. Once presenting the major components of Bitcoin, its basic characteristics and related concepts, in brief, our survey mainly focuses on the security and privacy aspects that can be found at various stages in the Bitcoin system, starting from transaction creation to its successful addition in the blockchain. We studied and emphasize the issue of user privacy and anonymity in this rapidly growing e-commerce industry. With the set of future research directions and open questions that we have raised, we hope that our work will motivate fledgling researchers towards tackling the security and privacy issues of Bitcoin systems.

\balance
\bibliographystyle{IEEEtran}
\bibliography{bare_jrnl}

\end{document}